\documentclass[aps,reprint,superscriptaddress,citeautoscript,showkeys,preprintnumbers,amsmath,amssymb,floatfix,nofootinbib,prb,longbibliography]{revtex4-2}
\usepackage{graphicx}
\usepackage{xcolor}
\usepackage{hyperref}
\hypersetup{colorlinks=true, linkcolor=blue, citecolor=blue, urlcolor=blue}
\usepackage{txfonts}
\usepackage[font=normalsize, caption=false]{subfig}
\usepackage{orcidlink}
\usepackage{makecell}
\newcommand{\tc}{\ensuremath{T_\text{c}}}

\newcommand{\omlog}{\ensuremath{\omega_{\log}}}

\begin{document}

\title{Anharmonic lattice dynamics and superconductivity in strained bulk and surface niobium} 

\author{Mihir Ranjan Sahoo \orcidlink{0000-0002-5528-7657}}
\affiliation{Institute of Theoretical and Computational Physics,
Graz University of Technology, Graz 8010, Austria}

\author{Roman Lucrezi \orcidlink{0000-0002-3117-3735}}
\affiliation{Department of Chemistry,
Stockholm University, SE-10691 Stockholm, Sweden}

\author{Pedro Nunes Ferreira \orcidlink{0000-0002-1135-0570}}
\affiliation{Institute of Theoretical and Computational Physics,
Graz University of Technology, Graz 8010, Austria}

\author{Chia-Nien Tsai \orcidlink{0009-0000-9107-4790}}
\affiliation{Smead Department of Aerospace Engineering
Sciences, University of Colorado, Boulder, 80303 CO, USA}

\author{Matthew Julian \orcidlink{0000-0003-3643-3484}}
\affiliation{Enterprise Science Fund, Intellectual Ventures, Bellevue, Washington 98005, USA}

\author{Rohit P. Prasankumar
\orcidlink{0000-0003-0902-2831}}
\affiliation{Enterprise Science Fund, Intellectual Ventures, Bellevue, Washington 98005, USA}

\author{Mahmoud I. Hussein \orcidlink{0000-0002-6583-1425}}
\affiliation{Smead Department of Aerospace Engineering
Sciences, University of Colorado, Boulder, 80303 CO, USA}
\affiliation{Department of Physics, University of Colorado, Boulder, 80302 CO, USA}

\author{Christoph Heil \orcidlink{0000-0001-9693-9183}}
\email{christoph.heil@tugraz.at}
\affiliation{Institute of Theoretical and Computational Physics,
Graz University of Technology, Graz 8010, Austria}

\date{\today}

\begin{abstract}
Using first-principles calculations, we investigate how homogeneous strain and crystallographic surface orientation modify the vibrational and superconducting properties of niobium. For bulk Nb, tensile strain strongly softens the phonon spectrum and enhances the electron--phonon coupling, increasing the superconducting transition temperature within the harmonic approximation from 9.5\,K at equilibrium to 14.6\,K at $\sim\!6\%$ lattice expansion. Anharmonic corrections are small near equilibrium but become increasingly significant under tensile strain and considerably reduce this enhancement at large expansion. For ultrathin Nb slabs with low-index (001), (110), and (111) surface orientations, harmonic phonon calculations exhibit imaginary modes, showing that anharmonic lattice effects are essential. 
To treat these effects efficiently, we train Nb-specific machine-learning interatomic potentials on bulk and slab first-principles configurations and use them to accelerate stochastic self-consistent harmonic approximation calculations, thereby obtaining anharmonically renormalized phonon modes that are combined with density-functional perturbation theory electron--phonon matrix elements to construct the Eliashberg spectral function. Among the three clean free-standing slabs examined, Nb(001) exhibits the strongest electron--phonon coupling and the highest calculated transition temperature of 10.1\,K, while Nb(110) and Nb(111) show progressively reduced pairing strength. Finally, by analyzing the Eliashberg spectral function and the functional derivative $\delta \tc{}/\delta\alpha^2F(\omega)$, we identify the phonon energy ranges most effective for superconducting pairing. Our results show that strain, surface termination, and anharmonic phonon renormalization provide complementary and interrelated microscopic routes for tuning superconductivity in Nb.
\end{abstract}


\maketitle

\section{\label{sec:level1}Introduction}

Enhancing the superconducting transition temperature \tc{} of conventional superconductors remains a longstanding goal in condensed-matter physics. Within the phonon-mediated mechanism, \tc{} is governed by a delicate interplay between electronic states at the Fermi level, phonon energy scales, and the strength and spectral distribution of the electron--phonon coupling (EPC). This makes superconductivity highly sensitive to perturbations that modify the electronic structure, the lattice dynamics, or both. Historically, approaches ranging from chemical substitution~\cite{ferreira2024,chemical-substitute-PRB} and the proximity effect~\cite{proximity-PRB, PhysRevMaterials.8.074801} to pressure~\cite{Ashcroft1968, Cs-Yt-pressure} and reduced dimensionality~\cite{shalnikov1938superconducting,blatt1963shape} have been used to influence superconducting behavior. Here, we focus on homogeneous strain~\cite{ekin1980strain} and surface-induced dimensional confinement~\cite{doi:10.1126/science.1106675,PhysRevLett.96.027005} as two routes for reshaping phonon spectra, EPC, and superconducting properties.

Strain is a particularly appealing tuning parameter because it modifies interatomic distances and force constants without necessarily changing the chemical composition. In conventional superconductors, lattice expansion or compression can alter bandwidths, the density of states at the Fermi level, phonon frequencies, and hence the EPC. Experimentally, however, strain is often difficult to vary as a clean and continuous control parameter. In thin films, it is commonly introduced through interaction with a substrate, arising from lattice mismatch or differential thermal contraction, and is therefore entangled with strain relaxation, interfacial bonding, defects, and surface morphology~\cite{Clavero2011NbStrain}. First-principles calculations provide a complementary, well-defined framework in which lattice parameters can be varied systematically while directly tracking the resulting changes in electronic structure, phonon spectra, EPC, and \tc{}.

Reduced dimensionality provides a second, closely related means of modifying superconductivity. In the thin-film limit, \tc{} is governed by a competition among quantum confinement, surface and interface electronic states, modified lattice dynamics, and reduced phase stiffness. Consequently, reducing the thickness of a superconducting film does not lead to a universal trend. Studies of monolayer and few-layer Pb films on Si reveal oscillations of \tc{} with thickness and a strong suppression in the few-layer regime, whereas one-atomic-layer Pb and In films remain superconducting with transition temperatures that depend sensitively on substrate coverage and bonding~\cite{PhysRevLett.96.027005,10.1126/science.1170775,Zhang2010}. In contrast, epitaxial Al films can exhibit an enhanced \tc{} when approaching the two-dimensional limit~\cite{ivry2014universal,sciadv.adf5500}. These contrasting behaviors show that the effect of reduced dimensionality on superconductivity is governed not by thickness alone, but by material-specific changes in electronic structure, phonon spectra, and EPC.

While the thickness dependence of \tc{} has been investigated in systematic comparative studies~\cite{kodama1983superconducting,park1985tc,minhaj1994thickness,gubin2005dependence}, the role of crystallographic surface orientation or atomic termination at fixed thickness remains much less explored. This distinction is important because the outermost atomic layers can host surface-derived electronic states, surface phonons, and altered bonding environments, all of which may substantially modify the EPC. Previous first-principles studies have shown, for example, that surface phonons can suppress \tc{} in ultrathin Pb films, while surface states can make a sizable contribution to the superconducting gap structure in few-layer MgB$_2$~\cite{Noffsinger2010,Bekaert2017}. A termination-resolved comparison of otherwise similar ultrathin films therefore provides a controlled way to isolate how the atomic structure of the surface affects lattice dynamics, EPC, and superconductivity.

Niobium is particularly well suited for investigating these effects. It has the highest superconducting transition temperature among elemental superconductors at ambient pressure, \tc{} $\approx 9.2$~K, and is widely used in superconducting radio-frequency cavities, resonators, and quantum devices~\cite{Hamlin2015,Balachandran2021Microstructure,Casalbuoni2005SurfaceSC}. At the same time, Nb is known to be highly sensitive to microstructure~\cite{microstructure-Nb} and near-surface conditions~\cite{Nb-near-surface}. Experiments on Nb thin films have shown that lattice mismatch, grain structure, surface morphology, and crystallographic quality can influence superconducting properties~\cite{Clavero2011NbStrain,Odobesko2019}. Recent studies further emphasize that the atomic-scale structure of Nb surfaces can strongly affect EPC and superconducting behavior: for example, a metallic Nb(100) surface exhibits a reduced EPC relative to bulk Nb, while the oxidized $(3\times1)$-O/Nb(100) reconstruction shows an even smaller coupling~\cite{ThompsonNb100HAS2024,ThompsonONb100HAS2024}. 
STM/STS on clean Nb(110) single-crystal surfaces has found a superconducting gap 
consistent with the bulk value~\cite{Odobesko2019}, and surface preparation procedures 
have recently been established for Nb(111)~\cite{Goedecke_STM_2023}; angle-resolved photoemission spectroscopy (ARPES) has also been reported for epitaxial Nb(110) thin films~\cite{Xiang_in_2017}. However, these measurements were performed on individual orientations rather than as systematic comparisons, and an orientation-resolved study of superconducting properties across different Nb surface terminations in the ultrathin limit has not yet been reported experimentally, to the best of our knowledge.
A unified microscopic understanding of how homogeneous strain and crystallographic surface orientation modify the phonon spectrum, EPC, and \tc{} of Nb is still lacking.

A computational challenge arises from the surface lattice dynamics, which  introduce instabilities not encountered in harmonic calculations of bulk Nb. Reduced coordination and broken translational symmetry soften surface force constants and enhance anharmonic contributions, so harmonic phonon calculations for Nb thin films and slabs can yield imaginary modes. A reliable evaluation of EPC and superconductivity therefore requires an anharmonic treatment of the phonon spectrum~\cite{lucrezi2022silico,LucreziBaSiH8Quantum2023,Roman-LuH3,Eva-NbN,Eva-Zureck-PdCuH2,lu2026}. The stochastic self-consistent harmonic approximation (SSCHA) provides such a framework by variationally optimizing an effective harmonic Hamiltonian on the anharmonic Born--Oppenheimer energy surface~\cite{SCHA,SSCHA}. However, its stochastic formulation requires force evaluations for large ensembles of displaced configurations, often amounting to thousands of first-principles calculations. This cost makes a direct \textit{ab initio} SSCHA treatment impractical unless the potential-energy surface can be sampled efficiently using interatomic potentials~\cite{novikov2021mlip,lucrezi2022silico,LucreziBaSiH8Quantum2023,Roman-LuH3,Eva-Zureck-PdCuH2}. At the same time, the relevant configurations involve thermally displaced, low-coordination surface environments, which are not reliably represented in bulk-trained or generic transferable interatomic potentials. Surface-specific first-principles training data are therefore essential for extending SSCHA-based anharmonic lattice-dynamical calculations to Nb thin films and slabs.

In this work, we combine density functional theory (DFT), density functional perturbation theory (DFPT), surface-trained machine-learning interatomic potentials, SSCHA, and isotropic Migdal--Eliashberg (ME) theory to investigate superconductivity in strained bulk Nb and in free-standing ultrathin Nb slabs with low-index surface orientations (001), (110), and (111). We first establish bulk Nb as a reference system and use controlled variations of the lattice constant to quantify how tensile and compressive strain modify the electronic structure, phonon spectra, EPC, and \tc{}. We then construct Nb-specific machine-learning interatomic potentials trained on first-principles bulk and surface configurations, including displaced structures representative of anharmonic surface dynamics. These potentials enable efficient SSCHA calculations of anharmonically stabilized phonon spectra for the Nb slabs, which in turn are used to evaluate the resulting surface-sensitive EPC and superconducting properties. By comparing slabs of similar thickness but different crystallographic surface orientations, we isolate how surface termination modifies the lattice dynamics and superconducting properties. Finally, through an analysis of the Eliashberg spectral function and the functional derivative $\delta \tc{}/\delta\alpha^2F(\omega)$, we identify the phonon energy ranges that most effectively contribute to superconducting pairing. This provides a microscopic picture of how strain, surface orientation, and anharmonic lattice dynamics reshape phonon-mediated superconductivity in Nb.

\section{Computational methods}
\label{sec:methods}
All electronic-structure calculations were carried out within DFT~\cite{hohenberg1964,kohn1965} using the Quantum~\textsc{Espresso} package~\cite{QE_2017}, the PBE exchange-correlation functional~\cite{perdew1996}, and scalar-relativistic optimized norm-conserving Vanderbilt (ONCV) pseudopotentials~\cite{ONCV,pseudo-dojo}.
Surface slabs were constructed with three layers for the (001) and (110) terminations and two layers for the (111) termination, where a layer denotes one repeat unit of the conventional cubic cell along the surface normal, comprising two atomic planes for (001) and (110) and six for (111). The resulting slabs have thicknesses of approximately 8.5--11~\AA{} and are separated from their periodic images by a 20~\AA{} vacuum region. Full details of the numerical setup are provided in Appendix~\ref{app:compdetails}.
 
Harmonic phonon dispersions and EPC matrix elements were computed within DFPT~\cite{DFPT}. The Eliashberg spectral function was constructed from the DFPT phonon linewidths $\gamma_{\mathbf{q}\nu}$ as
\begin{equation} 
\alpha^2F(\omega) = \frac{1}{2\pi N(E_\text{F})} \sum_{\mathbf{q}\nu} \frac{\gamma_{\mathbf{q}\nu}}{\omega_{\mathbf{q}\nu}}
\delta(\omega - \omega_{\mathbf{q}\nu}),
\label{equation1}
\end{equation}
where $N(E_\text{F})$ is the electronic density of states at the Fermi level and $\omega_{\mathbf{q}\nu}$ is the frequency of phonon branch $\nu$ at wavevector $\mathbf{q}$. The total EPC constant $\lambda$ and the logarithmic average phonon frequency \omlog{} follow as 
\begin{equation}
\lambda = 2\int_0^{\infty} \frac{\alpha^2F(\omega)}{\omega}\,d\omega,
\end{equation}
\begin{equation}
\omlog{} = \exp\!\left[\frac{2}{\lambda}\int_0^{\infty} \frac{\alpha^2F(\omega)}{\omega}\ln\omega\,d\omega \right].
\end{equation}
To visualize the distribution of EPC strength across phonon branches and wavevectors, mode- and wavevector-resolved coupling matrix elements were additionally obtained via Wannier interpolation using the \textsc{EPW} code~\cite{ME-EPW,EPW} with maximally localized Wannier functions~\cite{mosfoti2008,marzari2012}. These quantities are displayed as fat bands, with the line thickness indicating the relative EPC strength, in Figs.~\ref{fig:fig2_bulk_strain_main}(d) and (e). Superconducting properties were obtained by solving the isotropic ME equations in the Fermi-surface-restricted approximation using the IsoME code~\cite{lucrezi2024,isome}. The Fermi-surface-averaged screened Coulomb interaction $\mu = W(E_\text{F},E_\text{F}) = 0.38$, taken from the first-principles GW calculation of Ref.~\cite{isome}, is used
throughout; this corresponds to $\mu^{*}_\text{AD} \approx 0.12$ in Allen--Dynes estimates, according to the Morel--Anderson relation~\cite{MorelAnderson1962}
\begin{equation}
\mu^{*} = \frac{\mu}{1+\mu\ln\left(E_\text{el}/\omega_\text{ph}\right)},
\label{eq:morel-anderson}
\end{equation}
where $E_\text{el}$ and $\omega_\text{ph}$ denote characteristic electronic and phonon energy scales, respectively.

To examine the sensitivity of \tc{} with respect to the Coulomb pseudopotential, we considered two approaches. First, $\mu$ was rescaled according to the change in the density of states at the Fermi level as follows:
\begin{equation}
\mu_{\mathrm{new}} = \mu_{\mathrm{bulk}}
\frac{N(E_F)_{\mathrm{new}}}{N(E_F)_{\mathrm{bulk}}},
\end{equation}
using $\mu_{\mathrm{bulk}} = 0.38$. Second, for all systems, \tc{} was additionally evaluated using $\mu^*_{\mathrm{AD}} = 0.10$ and $0.16$, compared with the reference value of $0.12$. In addition, for the slabs, $N(E_F)$ was normalized to two Nb atoms, allowing comparison with the two-atom bulk cell. The results are summarized in Table~\ref{tab:mu-sensitivity} of Appendix~\ref{app:addfigures}.
 
\begin{figure}[t]
    \centering
    \includegraphics[width=\linewidth]{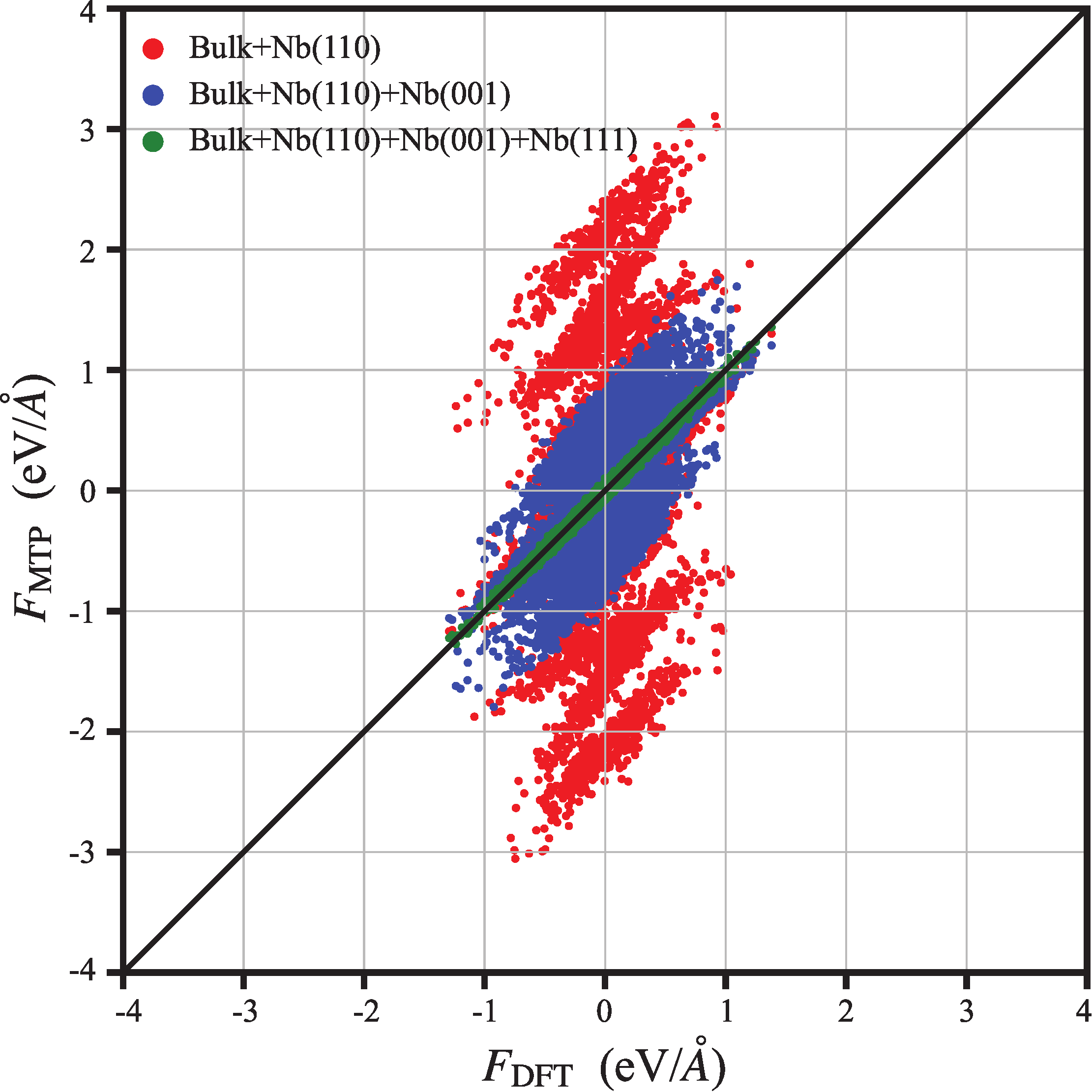}
    \caption{Parity plot of MLIP-predicted versus DFT atomic forces for Nb bulk and surface configurations. Each color corresponds to an MLIP trained on a progressively expanded dataset: Bulk$+$Nb(110) (red), Bulk$+$Nb(110)$+$Nb(001) (blue), and Bulk$+$Nb(110)$+$Nb(001)$+$Nb(111) (green). The diagonal line indicates perfect agreement.}
    \label{fig:fig1_force-plain}
\end{figure}

Harmonic phonon calculations for all Nb surface terminations yield imaginary modes, as discussed below, indicating that anharmonic lattice effects are essential for a physically meaningful description. We treat these using the SSCHA~\cite{SCHA,SSCHA}, which variationally optimizes an effective harmonic Hamiltonian on the anharmonic Born--Oppenheimer energy surface. In addition, to assess the role of anharmonicity under strain, we perform SSCHA calculations for bulk Nb at all lattice constants considered. Because the SSCHA requires forces for large stochastic ensembles of displaced configurations, direct first-principles sampling would be computationally prohibitive for the slab geometries considered here. We therefore train Nb-specific machine-learning interatomic potentials (MLIPs)~\cite{novikov2021mlip} on DFT bulk and surface configurations, following the workflow established in Refs.~\cite{lucrezi2022silico,LucreziBaSiH8Quantum2023,Roman-LuH3,Eva-Zureck-PdCuH2}. A key aspect of the training is the explicit inclusion of configurations from all three surface terminations, with atomic displacements sampled from the SSCHA stochastic ensemble, ensuring that the MLIP covers the anharmonic regions of the potential-energy surface relevant for surface dynamics. The accuracy of the trained potential is validated by comparing MLIP and DFT forces and energies on independent test configurations for each termination; results are shown in Fig.~\ref{fig:fig1_force-plain} and Appendix~\ref{app:addfigures}.

For the slab EPC and the anharmonic bulk results, anharmonic phonon frequencies obtained from the SSCHA free-energy Hessian replace the harmonic ones in Eq.~(1), while the electronic part of the electron--phonon interaction, which determines the phonon linewidths $\gamma_{\mathbf{q}\nu}$, is evaluated around the DFT-relaxed atomic structure, following Refs.~\cite{lucrezi2022silico,LucreziBaSiH8Quantum2023,Roman-LuH3}. This approximation is expected to be reasonable when the SSCHA centroid positions remain close to the DFT-relaxed structure, as discussed for all three slabs in Sec.~\ref{subsec:nb-surf}.

\section{Results and Discussion}
\subsection{Bulk Nb}
\label{subsec:bulk-nb}
\begin{figure*}[t]
    \centering
    \includegraphics[width=1.0\linewidth]{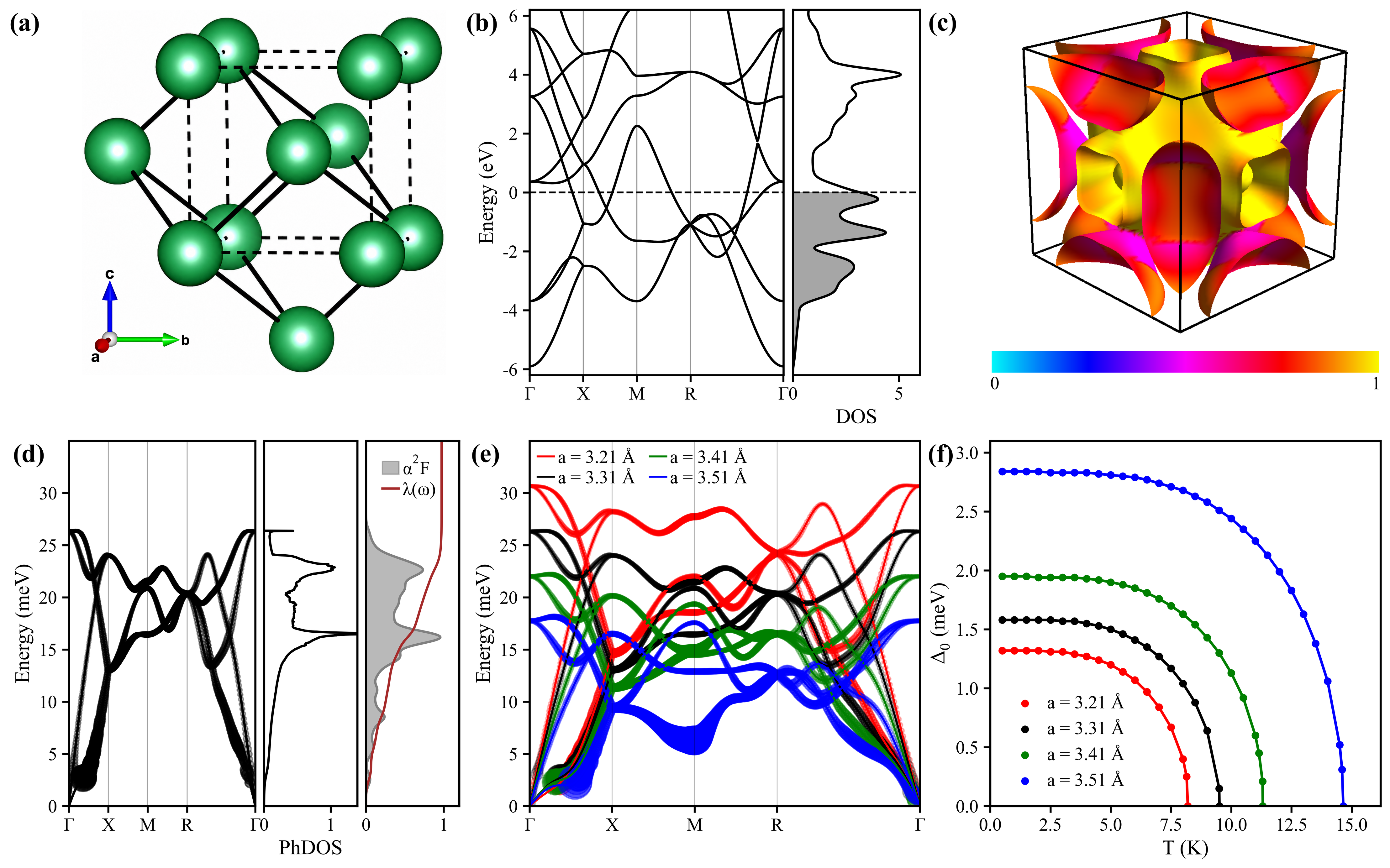}
    \caption{(a) Crystal structure of bulk Nb; dashed and solid lines indicate the conventional and primitive unit cells, respectively. (b) Electronic band structure along high-symmetry directions with the total DOS (states/eV) shown on the right; the dashed horizontal line marks the Fermi level ($E = 0$). (c) Nb $4d$ orbital-projected Fermi surface colored by the $4d$ orbital weight (color scale shown below). (d) Harmonic phonon dispersion with the phonon DOS, Eliashberg spectral function $\alpha^2F(\omega)$, and cumulative EPC constant $\lambda(\omega)$. The line thickness in the dispersion is proportional to the mode-resolved EPC strength. (e) Harmonic phonon dispersions at four lattice constants ($a = 3.31$\,\AA\ being the equilibrium value) with mode-resolved EPC shown as fat bands. (f) Superconducting gap $\Delta_0$ as a function of temperature obtained from isotropic ME calculations for each lattice constant; dots are calculated values, lines are guides to the eye.}
    \label{fig:fig2_bulk_strain_main}
\end{figure*}

We begin by summarizing the electronic, vibrational, and superconducting properties of bulk Nb as a reference. Elemental Nb crystallizes in the body-centered cubic structure (space group Im$\bar{3}$m). Our optimized lattice constant, $a=3.31$\,\AA, agrees well with the experimental value of $a=3.3004$\,\AA~\cite{StraumanisNb1970,RobergeNb1975}. The electronic band structure is shown in Fig.~\ref{fig:fig2_bulk_strain_main}(b), and the corresponding Fermi surface, dominated by Nb $4d$ states, is shown in Fig.~\ref{fig:fig2_bulk_strain_main}(c). Figure~\ref{fig:fig2_bulk_strain_main}(d) reports the harmonic phonon dispersion, phonon DOS, and Eliashberg spectral function $\alpha^2F(\omega)$. The calculated phonon energy scale and dispersion are consistent with x-ray and neutron-scattering measurements, while the magnitude of $\lambda$ and the structure of $\alpha^2F(\omega)$ agree with the established tunneling and neutron-scattering literature on Nb~\cite{ShapiroNbEPI1975,BostockNb1976,ButlerNbLinewidth1977,HoltNbPhonons2002}. Comparing the phonon DOS and $\alpha^2F(\omega)$ shows that the EPC is distributed over a broad range of phonon modes, with two prominent maxima around 17\,meV and 23\,meV associated with relatively flat phonon branches. From isotropic ME calculations~\cite{isome}, we obtain a total EPC constant $\lambda \approx 1.0$, a superconducting critical temperature $\tc{}\approx9.5$~K, and a zero-temperature superconducting gap $\Delta_0=1.6$\,meV, all in good agreement with experimental values~\cite{PRL-Nb,IKUSHIMA1969873,Park-Nb}.

To investigate the effect of lattice deformation, we systematically vary the lattice constant of bulk Nb from 3.21\,\AA\ to 3.51\,\AA, corresponding to isotropic (hydrostatic) compressive ($-3\%$) and tensile ($+3\%$ and $+6\%$) strains relative to the equilibrium value of 3.31\,\AA, and compute the phonon dispersions with mode-resolved EPC (Fig.~\ref{fig:fig2_bulk_strain_main}(e)). Such strain effects have been observed experimentally in epitaxial Nb films, where lattice mismatch with the substrate influences superconducting properties~\cite{Clavero2011NbStrain}. By again solving the isotropic ME equations for each strained structure with harmonic phonons, we find that compressive strain slightly suppresses \tc{}, whereas tensile strain significantly enhances it, reaching $\sim$14.6\,K at the largest expansion,  as shown in Fig.~\ref{fig:fig2_bulk_strain_main}(f). The superconducting gap follows this trend, reaching $\Delta_0 \approx 2.8$\,meV for the largest lattice parameter.

This enhancement is primarily driven by a strong increase in the EPC constant, which rises from $\lambda \approx 1.0$ at equilibrium to $\lambda \approx 1.8$ at $+6\%$ strain. According to the simplified Hopfield expression~\cite{Hopfield}, $\lambda \propto N(E_\text{F})\langle I^2 \rangle / (M\langle \omega^2 \rangle)$, the softening of phonon frequencies under tensile strain leads to a significant increase in coupling strength due to its inverse dependence on $\langle \omega^2 \rangle$. Although the characteristic phonon frequency scale given by \omlog{} decreases with increasing lattice constant, the enhancement in $\lambda$ dominates, resulting in an overall increase in \tc{}. In addition, tensile strain induces a slight increase in the density of states (Nb 4$d$ states) at the Fermi level due to band narrowing (see Fig.~\ref{fig:fig7_bulk_elbands_dos} in Appendix~\ref{app:addfigures}), which provides a secondary contribution to the enhancement of EPC. The calculated superconducting parameters are summarized in Table~\ref{tab:anharmonic-bulk}.  These results are robust with respect to the choice of Coulomb pseudopotential. Rescaling $\mu$ with the strain-dependent $N(E_F)$ changes \tc{} by at most 0.3~K, while varying $\mu^*_{\mathrm{AD}}$ between 0.10 and 0.16 preserves the increase of \tc{} with tensile strain (Table~\ref{tab:mu-sensitivity} in Appendix~\ref{app:addfigures}).

\begin{table*}[hbtp]
\centering
\caption{Harmonic and anharmonic superconducting properties of bulk Nb at different lattice parameters. The anharmonic results are obtained from SSCHA calculations at 5~K. The electron--phonon coupling constant ($\lambda$), logarithmic average phonon frequency ($\omega_{\log}$), zero-temperature superconducting gap ($\Delta_0$), and superconducting transition temperature (\tc{}) are reported for both calculations.}
\label{tab:anharmonic-bulk}

\small
\setlength{\tabcolsep}{4pt}

\begin{tabular}{c c c c c c c c c c}
\hline\hline
$a$ (\AA) &
$\epsilon$ &
$\lambda_{\mathrm{har}}$ &
$\lambda_{\mathrm{anh}}$ &
\makecell{$\omega_{\log}^{\mathrm{har}}$\\(meV)} & 
\makecell{$\omega_{\log}^{\mathrm{anh}}$\\(meV)} & 
\makecell{$\Delta_{0}^{\mathrm{har}}$\\(meV)} & 
\makecell{$\Delta_{0}^{\mathrm{anh}}$\\(meV)} & 
\makecell{$T_\text{c}^{\mathrm{har}}$\\(K)} & 
\makecell{$T_\text{c}^{\mathrm{anh}}$\\(K)} \\
\hline

3.21 & $-3\%$ & 0.81 & 0.74 & 16.50 & 15.60 &1.32 &0.99  & 8.2  & 6.2 \\

3.31 & $0\%$ & 0.97 & 0.87 & 12.80 & 14.43 &1.58 &1.40   & 9.5  & 8.5 \\

3.41 & $+3\%$ & 1.11 & 0.98 & 12.05 & 12.78 &1.95 &1.62 & 11.3 & 9.7 \\

3.51 & $+6\%$ & 1.76 & 0.90 & 8.56  & 12.67 &2.84 &1.34  & 14.6 & 8.1 \\

\hline\hline
\end{tabular}
\end{table*}

Bulk Nb is dynamically stable at all four lattice constants considered in this work, and the harmonic results at equilibrium agree well with the experimental superconducting properties. However, tensile strain softens the phonon modes, potentially increasing the importance of anharmonic corrections. To assess this effect, we additionally calculated the phonon spectra of strained bulk Nb within SSCHA at 5~K (Fig.~\ref{fig:bulk-phon-har-anhar} in Appendix~\ref{app:addfigures}). The resulting superconducting parameters are listed in Table~\ref{tab:anharmonic-bulk} alongside the harmonic ones.

At equilibrium, the harmonic and anharmonic spectra are similar, with the corresponding \tc{} values of 9.5~K and 8.5~K, respectively. The differences become more pronounced with increasing strain. Although the enhancement of \tc{} persists at moderate tensile strain, the large increase predicted at $+6\%$ strain within the harmonic approximation is considerably reduced when anharmonic effects are included. We note that SSCHA calculations at 10 and 15~K yield phonon dispersions nearly identical to those obtained at 5~K as shown in Fig.~\ref{fig:bulk-phon-temp} in Appendix~\ref{app:addfigures}. This indicates that the anharmonic renormalization is dominated by zero-point fluctuations: at 15~K, $k_BT\approx1.3$~meV, with  thermal occupation increasing the mean-square displacements of phonons above 5~meV by less than 5\% relative to their zero-point values.

While the suppression of \tc{} under compressive strain is qualitatively consistent with experimental observations on epitaxially strained Nb films~\cite{Clavero2011NbStrain}, the predicted enhancement under tensile strain has not yet been directly verified for elemental Nb, to the best of our knowledge. Epitaxial growth on substrates with lattice constants exceeding that of bulk Nb could in principle impose tensile strain and provide a direct test of the predicted \tc{} enhancement. Although hydrostatic compression can be achieved using various high-pressure techniques, such as piston--cylinder~\cite{pistoncell}, diamond-anvil~\cite{Diamond,Diamond1}, cubic-anvil~\cite{cubic-anvil}, and clamped-cube pressure cells~\cite{cubic-clamped}, experimentally realizing homogeneous hydrostatic tension in bulk Nb remains challenging. At present, there is no straightforward experimental technique for applying uniform hydrostatic tension to a bulk Nb sample as far as we know. However, several indirect approaches can induce lattice expansion or tensile strain. These include (i) exploiting thermal-expansion mismatch between a sample and its substrate, which can generate tensile strain upon cooling~\cite{thermal-expansion}; (ii) epitaxial growth of thin films on substrates with larger in-plane lattice constants~\cite{lattice-expand,Clavero2011NbStrain}, thereby inducing biaxial tensile strain in the film plane; and (iii) chemical substitution or hydrogen intercalation~\cite{KOMANDER202163,intercalation}, which can modify the lattice parameters. Nevertheless, these approaches generally produce biaxial or anisotropic strain, or introduce additional chemical and electronic effects. The isotropic lattice expansion considered here should therefore be regarded as an idealized setting that isolates the effect of lattice expansion on the phonon spectrum, EPC, and \tc{}. Our calculations predict an increase of \tc{} under moderate lattice expansion and estimate its magnitude, with the harmonic results providing an upper estimate. These overall findings serve as a theoretical benchmark for future experimental investigations.

To further identify which phonon modes are most effective in enhancing superconductivity, we analyze the Eliashberg spectral function together with the functional derivative $\delta \tc{} / \delta \alpha^2F(\omega)$, following the formalism introduced by Bergmann and Rainer~\cite{Bergmann1973}. 
Within Eliashberg theory, a variation in the spectral function leads to a corresponding change in the transition temperature given by~\cite{MITROVI2002,functional-Mesa}
\begin{equation}
\Delta \tc{} = \int_{0}^{\infty}\!\text{d}\omega \, \frac{\delta \tc{}}{\delta \alpha^2F(\omega)} \, \Delta \alpha^2F(\omega).
\end{equation}
Previous studies have shown that this quantity typically exhibits a broad maximum near $\omega \approx 7k_B \tc{}$, which represents an optimal phonon energy scale for enhancing superconductivity ~\cite{Bergmann1973,functional-carbotte,functional-Mesa}. By comparing this characteristic scale with \omlog{}, as well as with the positions of the maxima of $\alpha^2F(\omega)$ and $\delta \tc{}/\delta \alpha^2F(\omega)$, one can assess how efficiently the phonon spectrum promotes superconductivity.

\begin{figure}[t]
    \centering
    \includegraphics[width=\linewidth]{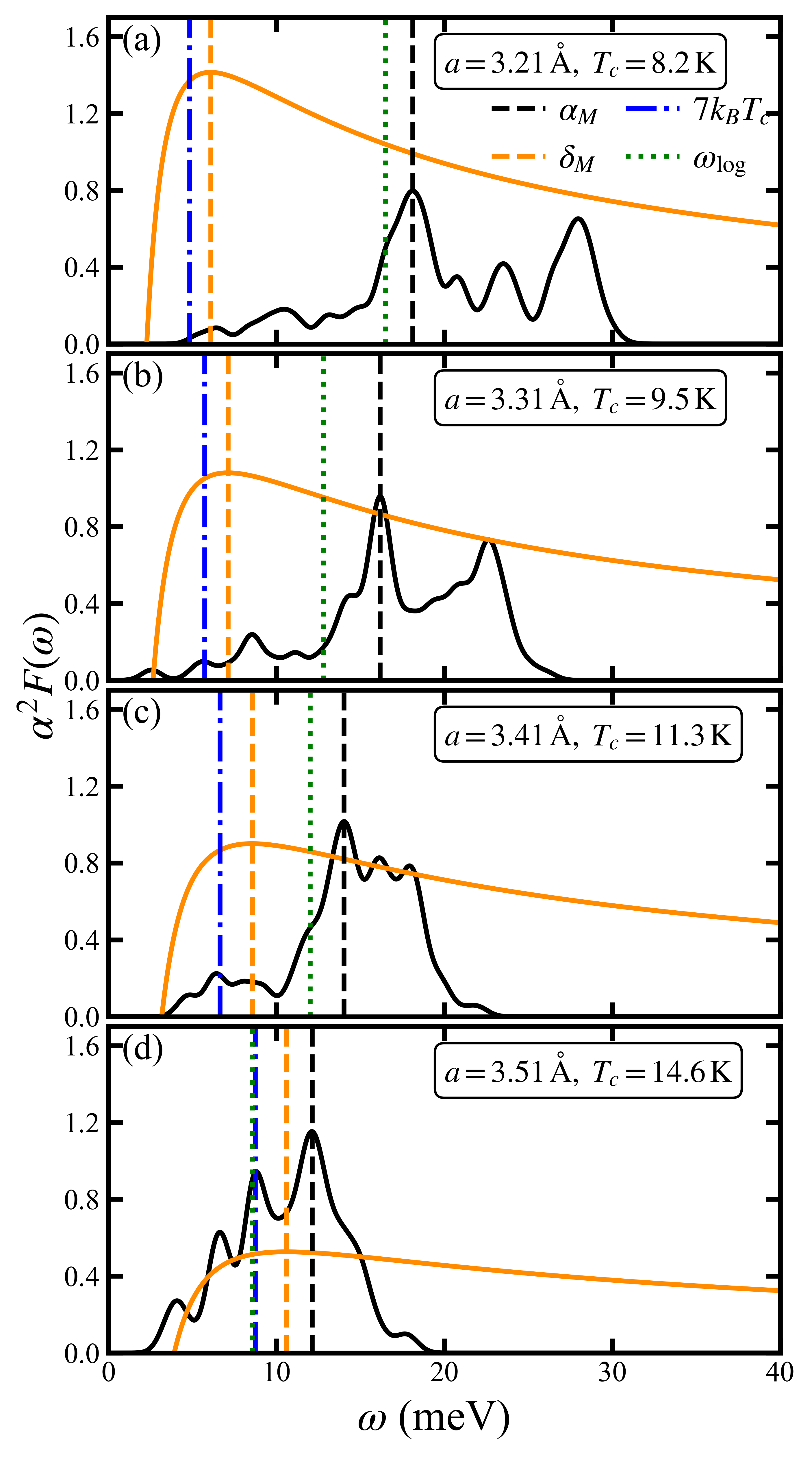}
    \caption{Eliashberg spectral function $\alpha^2F(\omega)$ (black) and functional derivative $\delta \tc{}/\delta\alpha^2F(\omega)$ (orange) for bulk Nb at four lattice constants (panels a--d). Vertical lines mark characteristic energy scales: the peak positions of $\alpha^2F(\omega)$ ($\alpha_M$, black dashed) and $\delta \tc{}/\delta\alpha^2F(\omega)$ ($\delta_M$, orange dashed), the optimal pairing scale $7k_B \tc{}$ (blue dash-dotted), and the logarithmic average phonon frequency \omlog{} (green dotted). The lattice constant and calculated \tc{} for each case are given in the panel insets.} 
    \label{fig:fig3_a2Fder_bulk}
\end{figure}

Figure~\ref{fig:fig3_a2Fder_bulk} shows $\alpha^2F(\omega)$ together with $\delta \tc{}/\delta\alpha^2F(\omega)$ for bulk Nb at different lattice parameters, obtained with harmonic phonons. Here, superconductivity in Nb is primarily governed by low- to intermediate-frequency phonons in the range of approximately $5$--$15$~meV. With increasing lattice parameter, the characteristic phonon energy scales soften systematically. In particular, the separation between the optimal pairing scale ($\sim 7k_B \tc{}$) and both the dominant peak of $\alpha^2F(\omega)$ ($\alpha_M$) and the logarithmic frequency \omlog{} decreases with increasing \tc, indicating that the phonon spectral weight shifts toward the optimal energy range for pairing. At the same time, the maximum of the functional derivative ($\delta_M$) shifts toward higher phonon energies, showing that phonons in this intermediate energy range increasingly contribute to superconductivity under tensile strain. 

These bulk results show that tensile strain is an effective tuning parameter for enhancing superconductivity in Nb: within the harmonic approximation, \tc{} rises by more than 50\% relative to its equilibrium value through phonon softening and the concomitant increase in EPC strength, and although anharmonic effects reduce the attainable enhancement at large strain, an increase of \tc{} persists at moderate tensile strain (Table~\ref{tab:anharmonic-bulk}). The functional-derivative analysis confirms that this enhancement is spectrally coherent at the harmonic level, i.e., the phonon weight shifts progressively toward the most effective pairing frequencies rather than simply redistributing uniformly.

\subsection{Nb Surfaces}
\label{subsec:nb-surf}
\begin{figure*}[t]
\centering
\includegraphics[width=\textwidth]{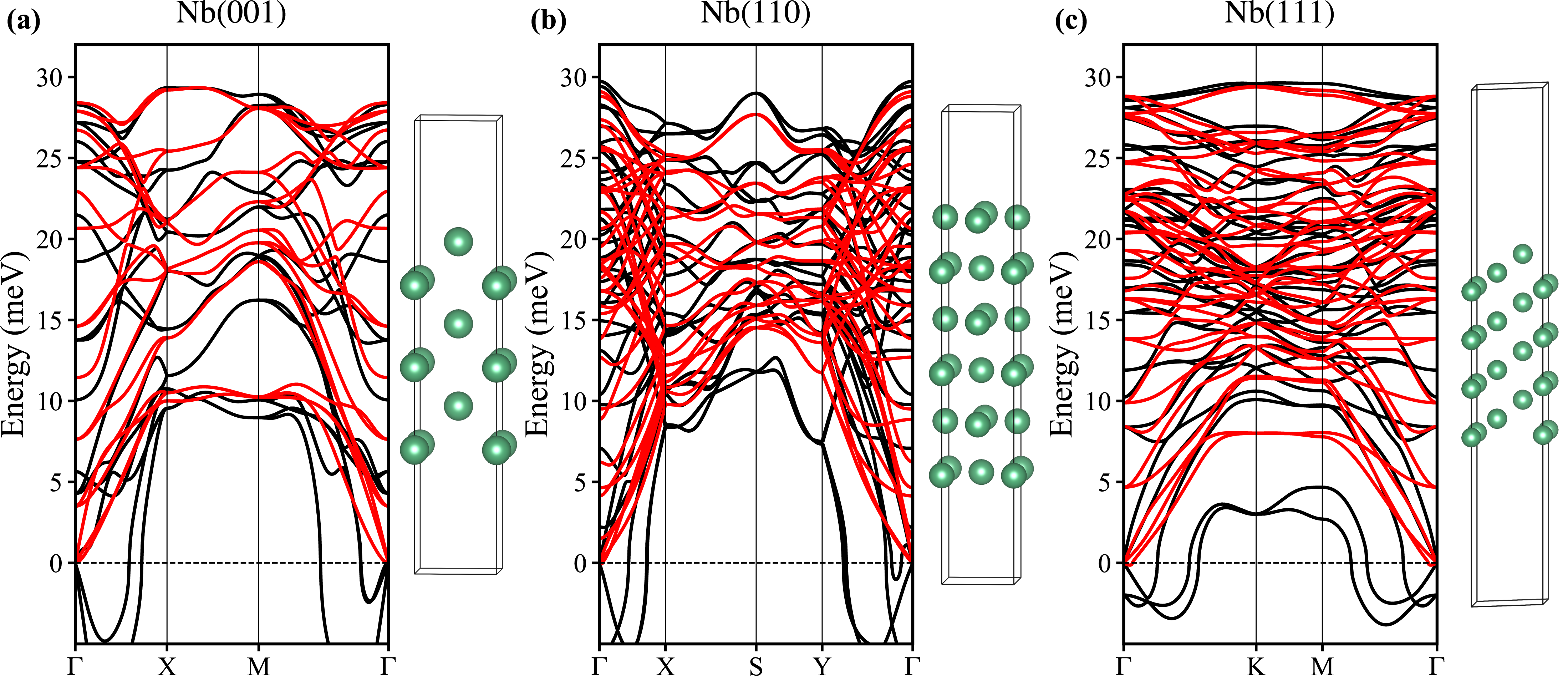}
\caption{Side-view atomic structures and corresponding phonon dispersions of Nb slabs: (a) three-layer Nb(001), (b) three-layer Nb(110), and (c) two-layer Nb(111) (see Sec.~\ref{sec:methods} for the definition of a layer). In each panel, the left subfigure shows the phonon dispersion along high-symmetry directions, while the right subfigure presents the atomic slab structure. Black lines denote harmonic phonon spectra, while red lines represent anharmonic phonon dispersions obtained from SSCHA calculations at 5~K in $2\times2\times1$ supercells; the same temperature and supercell size were used for all three terminations.}
\label{fig:fig4_surface_anharm}
\end{figure*}

With the bulk reference cases established, we turn to the surface systems, where reduced coordination and broken translational symmetry introduce additional and more complex modifications to the lattice dynamics, and, by extension, the EPC properties. The structures with low-index Nb surfaces featuring (001), (110) and (111) terminations are shown schematically in Fig.~\ref{fig:fig4_surface_anharm}(a)--(c). The slabs -- also often referred to as membranes -- have thicknesses of approximately 8.5--11\,\AA, which is comparable to the lower end of experimentally investigated ultrathin Nb films~\cite{ZaytsevaNbUltrathin2020}. It is important to note, however, that the slabs considered in this work represent clean, free-standing, crystalline reference systems that are idealized and should not be interpreted as models of substrate-supported, capped, oxidized, or disordered films.
The electronic structures and Fermi surfaces of these slabs are shown in Fig.~\ref{fig:fig8_slabs_elbands_dos_FS} in Appendix~\ref{app:addfigures}, and the phonon dispersions are shown in Fig.~\ref{fig:fig4_surface_anharm}(a)-(c). Within the harmonic approximation, all our slab models exhibit pronounced imaginary phonon modes, indicating dynamical instabilities. A key source of these instabilities is the reduced atomic coordination at the surface, which weakens interatomic restoring forces and leads to soft vibrational modes that become unstable in the harmonic limit~\cite{pallikara2022}.

Treating these instabilities requires going beyond the harmonic approximation, and we do so using the MLIP-accelerated SSCHA workflow described in Sec.~\ref{sec:methods} and Refs.~\cite{lucrezi2022silico,LucreziBaSiH8Quantum2023,Roman-LuH3}. 
We find that achieving accurate force predictions across all surface environments is non-trivial: reliable results are obtained only when configurations from all three surface terminations are explicitly included in the training dataset. This is clearly seen in Fig.~\ref{fig:fig1_force-plain}, where the systematic reduction in scatter with each added termination demonstrates the importance of surface-specific training data; full validation plots for forces and energies are provided in Appendix~\ref{app:addfigures}.

\begin{table}[t]
\centering
\caption{Superconducting parameters of Nb slabs with different crystallographic terminations, including the EPC constant $\lambda$, logarithmic phonon frequency \omlog{},  superconducting gap $\Delta_0$, and transition temperature \tc{} obtained from isotropic Migdal–Eliashberg theory using the anharmonic SSCHA phonons at 5~K (superscript anh).}\par
\vspace*{0.75em}
\label{tab:sc_params_surf}
\begin{tabular}{c c r r r}
\hline\hline
\noalign{\vskip 3pt}
\shortstack{Surface\\[3pt] termination} &
\shortstack{$\lambda^{\mathrm{anh}}$\\[3pt] \phantom{(K)}} &
\shortstack{$\omega_{\log}^{\mathrm{anh}}$\\[3pt] (meV)} &
\shortstack{$\Delta_0^{\mathrm{anh}}$\\[3pt] (meV)} &
\shortstack{$T_\text{c}^{\mathrm{anh}}$\\[3pt] (K)} \\[4pt]
\hline\hline
001 & 1.06 & 11.90 & 1.7 & 10.1 \\
110 & 0.86 & 13.89 & 1.3 &  7.8 \\
111 & 0.77 & 14.00 & 1.0 &  6.2 \\[2pt]
\hline\hline
\end{tabular}
\end{table}

\begin{figure}[t]
    \centering
    \includegraphics[width=1\linewidth]{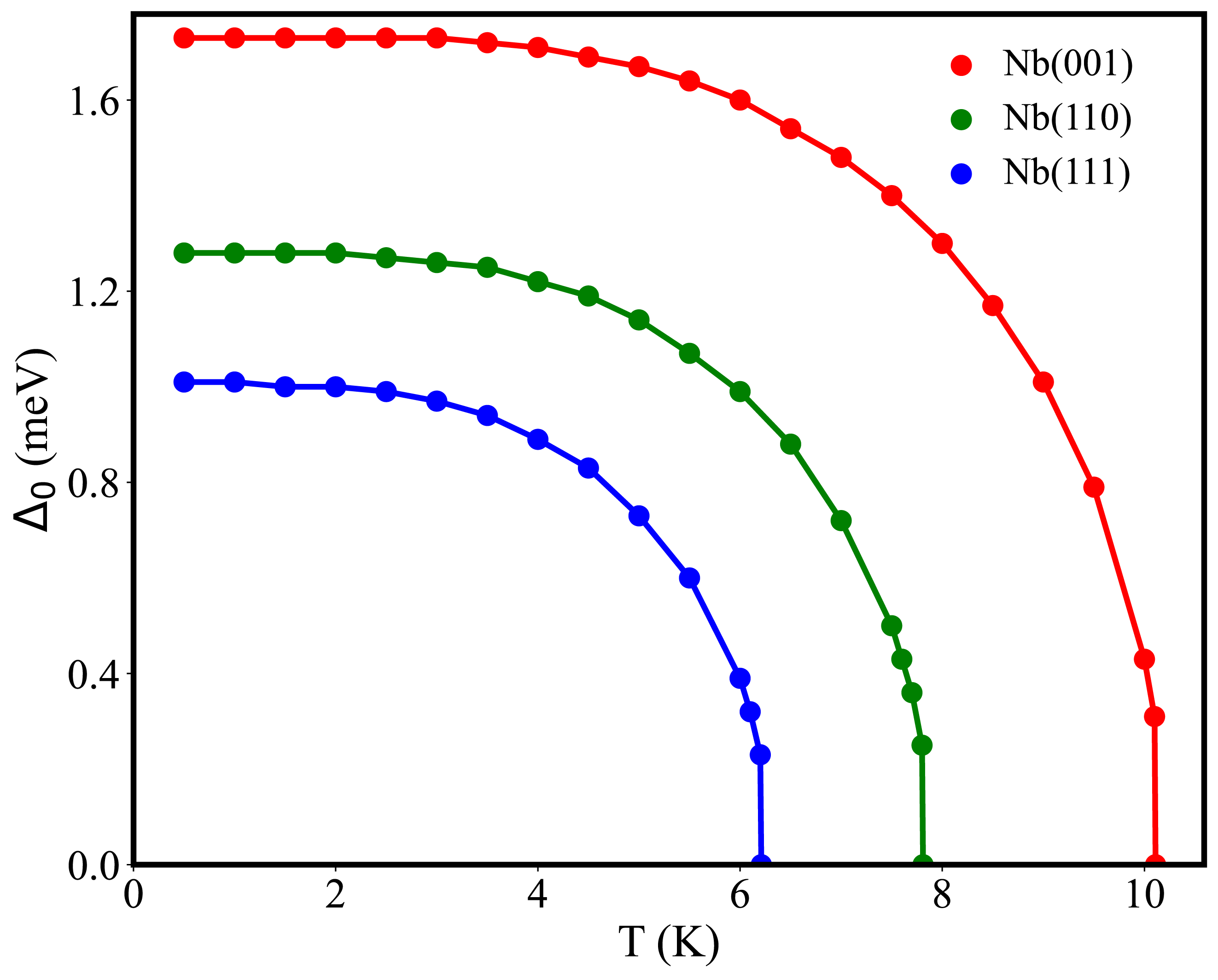}
    \caption{Temperature dependence of the superconducting gap $\Delta_0$ obtained from isotropic Migdal--Eliashberg calculations for the Nb(001), Nb(110), and Nb(111) slabs. Dots are calculated values; solid lines are guides to the eye.}  
    \label{fig:fig5_isoME_surf}
\end{figure}

Upon inclusion of anharmonic effects, we find that all soft modes are fully stabilized and shift to positive frequencies, indicating that anharmonic corrections are necessary for obtaining physically meaningful lattice dynamics results in the low-dimensional Nb models considered here. The anharmonic renormalization of phonon modes also directly influences the EPC, particularly through modifications of low-frequency vibrations that dominate superconducting pairing. The temperature dependence of the superconducting gap obtained from the ME calculations is shown in Fig.~\ref{fig:fig5_isoME_surf} for the three Nb surface terminations, and the corresponding superconducting properties are summarized in Table~\ref{tab:sc_params_surf}. 

Since the electron--phonon matrix elements used in Eq.~(\ref{equation1}) are evaluated at the DFT-relaxed geometry, whereas the phonon frequencies are obtained from the SSCHA, it is essential to assess the structural changes induced by anharmonicity. We therefore compare the DFT-relaxed atomic positions with those of the SSCHA centroid structures for all three terminations. For Nb(001) and Nb(111), the centroid shifts are very small. The root-mean-square (RMS) displacements are 0.005 and 0.010~\AA, respectively, while the maximum displacements are 0.008 and 0.019~\AA, respectively. These values are less than $1\%$ of the nearest-neighbor Nb--Nb distance ($\approx 2.87$~\AA). For Nb(110), the centroid shifts are relatively larger, with an RMS displacement of 0.092~\AA{} and a maximum displacement of 0.124~\AA, corresponding to about $4.3\%$ of the nearest-neighbor distance. These centroid shifts describe changes in the mean atomic structure and should be distinguished from the larger quantum and thermal fluctuations sampled within the SSCHA ensemble. For Nb(001) and Nb(111), the small shifts support the use of DFT-level electron--phonon matrix elements together with anharmonically renormalized phonons. For Nb(110), these relatively larger structural shifts may introduce some uncertainty, as a corresponding modification of the electron--phonon matrix elements cannot be excluded. This should be considered when interpreting the Nb(110) results.

For the finite-slab geometries studied, Nb(001) exhibits the highest superconducting \tc{} of 10.1\,K, while Nb(110) and Nb(111) show progressively reduced pairing strength. The ordering of the transition temperatures is also robust against changes in the Coulomb pseudopotential. Both rescaling $\mu$ according to $N(E_F)$ and varying $\mu^*_{\mathrm{AD}}$ between 0.10 and 0.16 preserve the ordering $T_\text{c}(001) > T_\text{c}(110) > T_\text{c}(111)$ (Table~\ref{tab:mu-sensitivity} in Appendix~\ref{app:addfigures}). This trend can be understood in terms of surface lattice dynamics: the (001) surface exhibits stronger phonon softening and anharmonic renormalization of low-frequency modes, which enhances the electron--phonon interaction ($\lambda = 1.06$), whereas the weaker coupling in the (110) and (111) surface systems reflects a reduced contribution from such soft vibrational modes.
These results highlight a clear orientation dependence of superconductivity within this set of comparable clean Nb slab geometries, consistent with experimental observations that surface modifications of lattice dynamics and EPC can influence superconducting behavior~\cite{ThompsonNb100HAS2024,ThompsonONb100HAS2024}. Furthermore, SSCHA calculations at 10~K yield phonon dispersions nearly identical to those obtained at 5~K for all three slab terminations (see Fig.~\ref{fig:surf-phon-temp} in Appendix~\ref{app:addfigures}).

It is instructive to compare these ideal-slab results with recent surface-sensitive helium atom scattering (HAS) measurements on Nb(100)~\cite{nb001note}. Thompson \textit{et al.} measured the surface EPC constant of a clean, unreconstructed metallic Nb(100) surface and obtained $\lambda_S=0.50\pm0.08$, approximately half of commonly reported bulk Nb values. Using this surface coupling together with a surface Debye temperature in McMillan-type estimates, they inferred a surface transition temperature in the range $\tc{}=1.4$--$3.6$~K~\cite{ThompsonNb100HAS2024}. A companion HAS study of the oxidized $(3\times1)$-O/Nb(100) reconstruction found an even smaller $\lambda_S=0.20\pm0.06$ and an estimated $\tc{}\leq 6.2\times10^{-3}$~K~\cite{ThompsonONb100HAS2024}. These values are substantially smaller than the Nb(001) value obtained here, but the comparison involves different physical quantities. HAS probes the near-surface electron density and vibrational response, with the strongest sensitivity to the outermost layers, whereas the present $\lambda$ and \tc{} are obtained from the Eliashberg spectral function of the complete finite slab and therefore incorporate contributions from all layers of the slab~\cite{MansonHASReview2022}.

This distinction has recently been put on a firmer microscopic footing by the \textit{ab initio} theory of elastic HAS developed by M\'endez \textit{et al.}~\cite{MendezHASPRL2025}. In that work, the temperature-dependent attenuation of elastic HAS intensities was linked directly to phonon-induced fluctuations of the surface electron density, and the theory was validated for both clean Nb(100) and the oxidized $(3\times1)$-O/Nb(100) reconstruction~\cite{McMillanNb100OxidePhonons2022}. Importantly, the resulting HAS response was shown to be a mode- and polarization-weighted surface quantity, with distinct contributions from bulk, resonant-surface, and true-surface phonons. Thus, the HAS-derived $\lambda_S$ and the superconducting $\lambda$ calculated here should be regarded as complementary measures of electron--phonon physics: the former characterizes the surface-weighted vibrational response relevant for He scattering, whereas the latter is the pairing interaction obtained from the full slab-averaged Eliashberg spectral function. This provides a natural explanation for the smaller experimentally inferred $\lambda_S$ while supporting the central conclusion that surface phonons play a decisive role in Nb slab and thin-film superconductivity.

\begin{figure}[hbtp]
    \centering
    \includegraphics[width=\columnwidth]{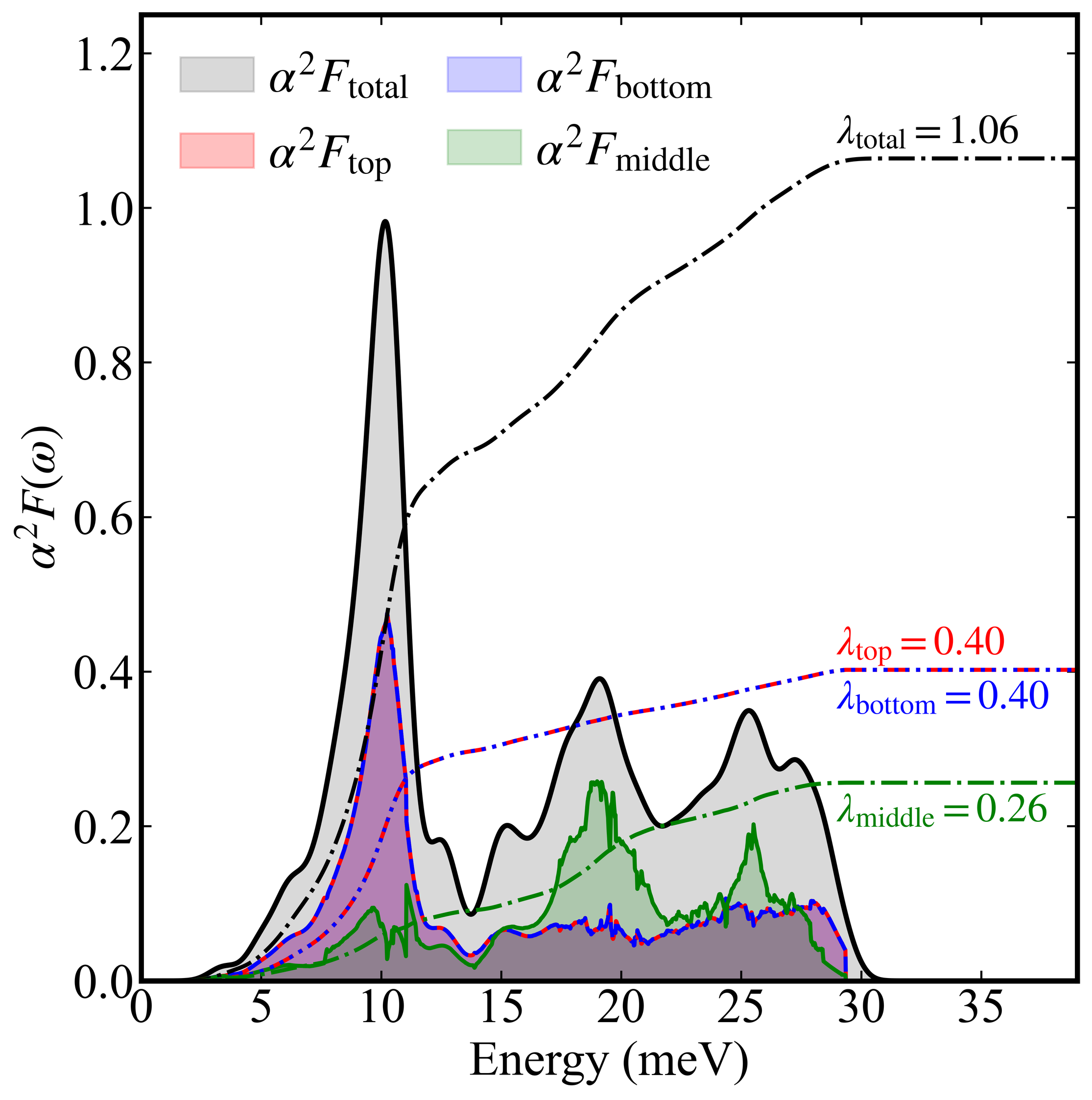}
    \caption{
    Layer-projected Eliashberg spectral function and cumulative 
    electron--phonon coupling for the three-layer Nb(001) slab. The shaded 
    regions show the layer-projected $\alpha^2F_i(\omega)$, while the 
    dash-dotted curves represent the corresponding cumulative 
    $\lambda_i(\omega)$. }
    \label{fig:layer-alpha2F}
\end{figure}

To examine where in the slab the coupling originates and to relate it to the HAS results, we decompose the Nb(001) slab Eliashberg spectral function into layer-resolved contributions. Assuming that the ratio of the Eliashberg spectral function to the phonon density of states is frequency-dependent but identical across all layers, we estimate the layer-projected Eliashberg function as
\begin{equation}
\alpha^2F_i(\omega)=
\alpha^2F_{\mathrm{total}}(\omega)
\frac{F_i(\omega)}{F_{\mathrm{total}}(\omega)},
\end{equation}
where $F_{\mathrm{total}}(\omega)$ is the total phonon density of states and $F_i(\omega)$ is the phonon density of states projected onto the atoms of layer $i$ (two atomic planes; see Sec.~\ref{sec:methods}), obtained using the SSCHA phonon eigenvectors. The corresponding layer-resolved EPC strength is calculated as
\begin{equation}
\lambda_i=2\int_0^{\omega_{\max}}
\frac{\alpha^2F_i(\omega)}{\omega}\,d\omega,
\end{equation}
where $\omega_{\max}$ is the maximum phonon frequency of the slab. For the three-layer Nb(001) slab, we obtain $\lambda_{\mathrm{top}}=\lambda_{\mathrm{bottom}}=0.40$ and $\lambda_{\mathrm{middle}}=0.26$, yielding a total coupling of $\lambda=1.06$. The equivalence of the top and bottom layer contributions is consistent with the symmetry of the slab. Since the $\lambda_i$ are additive contributions to the total coupling, they scale with the fraction of atoms in each layer and cannot be compared directly with an intrinsic surface coupling such as $\lambda_S$. Figure~\ref{fig:layer-alpha2F} further shows that the surface layers dominate the low-energy peak in $\alpha^2F(\omega)$ around 10~meV, which originates from soft, anharmonically renormalized surface modes. In contrast, the central layer contributes predominantly at higher energies, around 18--26~meV. Within this approximation, the outermost layers therefore couple more strongly than the interior of the slab, and the smaller $\lambda_S$ inferred from HAS cannot be attributed to a weaker electron--phonon coupling of the surface layers in our calculations. In fact, as discussed above, the HAS signal is a mode- and polarization-weighted response of the surface electron density rather than the pairing interaction within a layer entering the Eliashberg spectral function~\cite{MendezHASPRL2025}. Differences between the idealized free-standing slab and the experimental surface may also contribute to the discrepancy. These results also highlight the important role of surface-localized vibrational modes in the electron--phonon interaction of Nb(001).

\begin{figure}[t]
    \centering
    \includegraphics[width=1.0\linewidth]{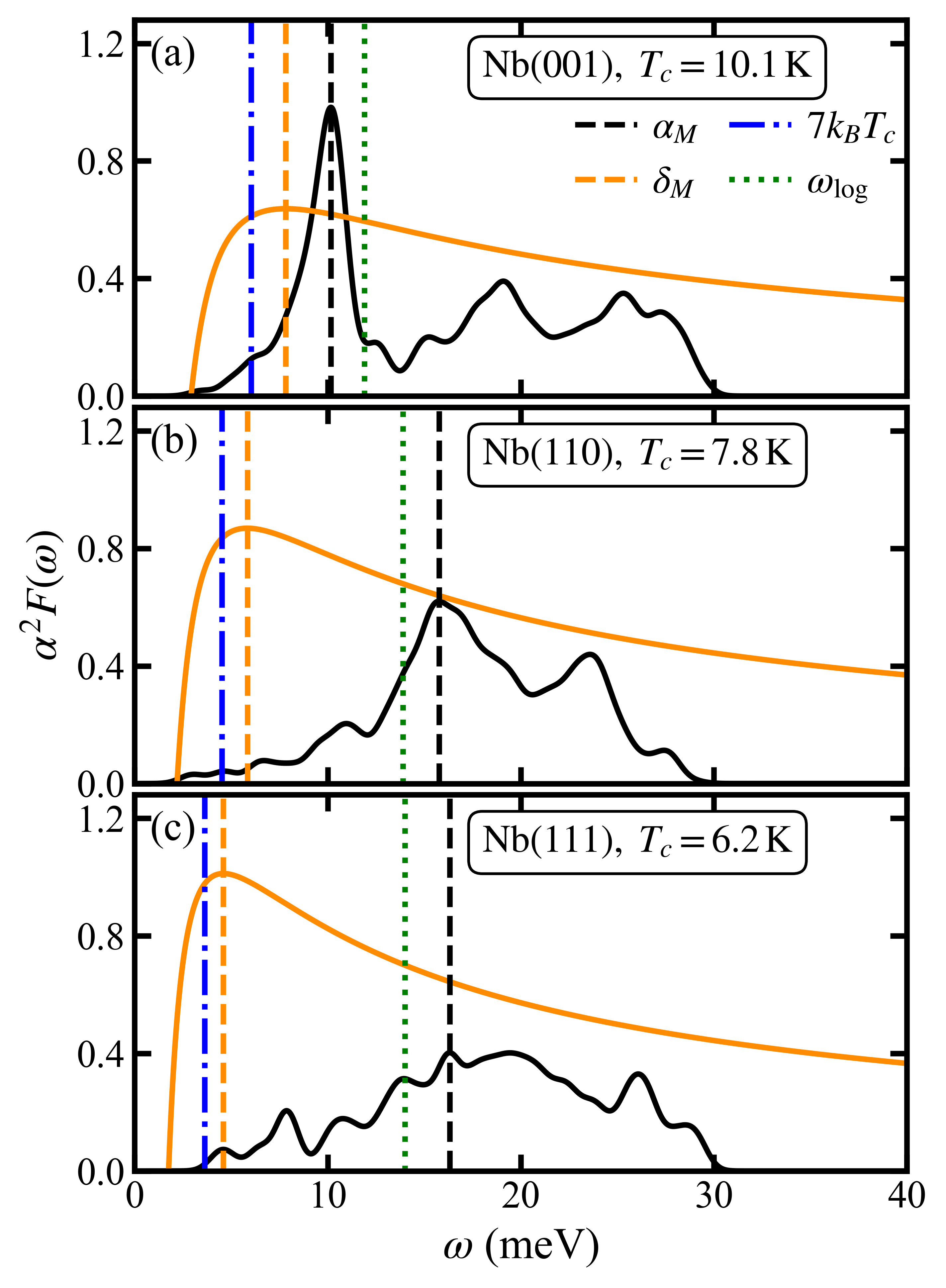}
    \caption{Eliashberg spectral function $\alpha^2F(\omega)$ (black) and functional derivative $\delta \tc{}/\delta\alpha^2F(\omega)$ (orange) for the three-layer Nb(001), three-layer Nb(110), and two-layer Nb(111) slabs (panels a--c) obtained using the 5~K SSCHA phonon spectra, computed in $2\times2\times1$ supercells for all three terminations. Vertical lines mark characteristic energy scales: the peak positions of $\alpha^2F(\omega)$ ($\alpha_M$, black dashed) and $\delta \tc{}/\delta\alpha^2F(\omega)$ ($\delta_M$, orange dashed), the optimal pairing scale $7k_B \tc{}$ (blue dash-dotted), and the logarithmic average phonon frequency \omlog{} (green dotted). The calculated \tc{} for each termination is given in the panel insets.} 
    \label{fig:fig6_a2Fder_slabs}
\end{figure}

For the finite-slab geometries considered in this work, the calculated orientation trend is also consistent with earlier first-principles slab calculations by Csire, Sch\"onecker, and \'Ujfalussy, who found that Nb(100) slabs can have transition temperatures above the bulk value because surface phonon softening and an enhanced McMillan--Hopfield parameter increase the EPC, whereas Nb(110) slabs exhibit weaker enhancement and lower transition temperatures~\cite{CsireNbSlabs2016}. R\"u{\ss}mann and Bl\"ugel later analyzed bulk Nb and Nb(110) surfaces within a density-functional Bogoliubov--de Gennes framework and emphasized that surface relaxations and phonon softening modify the local anomalous density, while the superconducting gap is averaged over the film thickness because the coherence length of Nb is much larger than the interlayer spacing~\cite{RussmannNb110BdG2022}. The present work extends these approaches by evaluating the EPC from the full Eliashberg spectral function and by anharmonically stabilizing the slab phonon spectra using SSCHA with surface-trained machine-learning interatomic potentials. This is essential because the harmonic slab phonons contain imaginary modes in our models, so a purely harmonic treatment would not provide a physically stable starting point for the EPC calculation.

An analysis of the functional derivative of \tc{} for the slab systems, shown in Fig.~\ref{fig:fig6_a2Fder_slabs}, reveals a consistent trend. 
Among the terminations considered, the Nb(001) surface, which exhibits the highest \tc{}, has the smallest separation between the optimal pairing scale ($\sim\!7k_B \tc{}$) and the characteristic phonon energies of the spectrum. In contrast, the Nb(111) surface shows the largest separation and correspondingly the lowest \tc{}, while the Nb(110) surface displays intermediate behavior. This trend again demonstrates that superconductivity in surface systems is governed by the alignment between the phonon spectral weight and the optimal pairing energy scale. When the dominant phonon modes lie closer to this energy range, the electron--phonon interaction becomes more effective in enhancing superconductivity, leading to higher \tc{}. 

These results mirror the bulk behavior and further highlight that the distribution of electron--phonon spectral weight plays a key role in optimizing \tc{}. 
In bulk Nb, superconductivity is governed by moderate EPC strength arising from intermediate-frequency phonon modes. Tensile strain enhances superconductivity through phonon softening and increased $\lambda$, whereas compressive strain suppresses it by stiffening the lattice. At the surface, reduced dimensionality and coordination introduce strong anharmonic lattice effects that must be treated beyond the harmonic approximation. The interplay between dimensionality, anharmonic phonon stabilization, and electron–phonon interaction ultimately determines whether superconductivity is enhanced or suppressed relative to the bulk. 
Additional calculations for thicker Nb(001) and Nb(111) slabs show that their transition temperatures evolve toward the bulk value with increasing thickness, while the ordering $T_\text{c}(001)>T_\text{c}(111)$ is preserved. Thus, the magnitude of the orientation dependence discussed above is characteristic of the ultrathin regime. The ordering of \tc{} does not follow the normalized $N(E_F)$ (largest for Nb(110); Table~\ref{tab:mu-sensitivity}), as it would if quantum-well states at the Fermi level dominated. Instead, it follows the softening of the surface modes, reflected in \omlog{}, indicating that the orientation dependence is driven primarily by the surface lattice dynamics. In-plane epitaxial strain on Nb(001) modifies \tc{} with a trend opposite to that of bulk Nb: \tc{} rises to 14.0~K at $-3\%$ and decreases to 6.7~K at $+6\%$ (Table~\ref{tab:surface-strain} in Appendix~\ref{app:strain-thick-slab}). Effects of carrier doping, e.g., through charge transfer at interfaces or alloying, are not considered here and remain a direction for future work.

These findings highlight the central role of lattice dynamics and anharmonicity in tuning superconductivity in reduced-dimensional metallic systems.

\section{Conclusions}

We have investigated how homogeneous strain and crystallographic surface orientation shape superconductivity in Nb by combining first-principles calculations of electronic structure and lattice dynamics with isotropic ME theory. For bulk Nb, within the harmonic approximation, tensile strain emerges as a highly effective lever: we find that phonon softening and the concomitant increase in EPC strength raise \tc{} from 9.5\,K at equilibrium to 14.6\,K at $+6\%$ expansion, a more than 50\% enhancement. Additional SSCHA calculations show that anharmonic effects, while small near equilibrium, become increasingly consequential under tensile strain and considerably reduce the enhancement at large strain, although an increase in \tc{} persists at moderate strain. 

For the Nb slabs, we demonstrate that the harmonic approximation fails for all three low-index terminations, and that anharmonically stabilized phonon spectra obtained from MLIP-accelerated SSCHA calculations are essential for a physically meaningful evaluation of the EPC. The resulting superconducting properties show a clear orientation dependence: Nb(001) exhibits the strongest EPC and the highest superconducting pairing scale, while Nb(110) and Nb(111) show progressively reduced pairing strength. Resolving the Nb(001) coupling by layer shows that the outermost layers dominate the soft-mode contribution and couple more strongly than the interior of the slab. Additional calculations for thicker slabs show that the transition temperatures of both terminations considered approach the bulk value with increasing thickness while the ordering of Nb(001) and Nb(111) is preserved. Thus, the magnitude of the orientation dependence identified here is characteristic of the ultrathin regime. Moreover, in-plane epitaxial strain modifies the superconductivity of Nb(001) with a trend opposite to that of bulk Nb. Compressive strain enhances \tc{}, while tensile strain suppresses it, and anharmonic effects remain important under strain. The orientation dependence identified here constitutes a concrete prediction that should be accessible to surface-sensitive probes such as tunneling spectroscopy or angle-resolved photoemission on well-characterized single-crystal Nb surfaces of different orientations, complementing the measurements that have so far been performed on individual terminations.

Across both the bulk strain and surface orientation studies, the functional derivative $\delta \tc{}/\delta\alpha^2F(\omega)$ provides a unified energy-resolved interpretation of the observed trends, showing that the spectral distribution of the EPC is as important as its total strength in determining \tc{}. Applied to Nb, this analysis identifies strain, surface orientation, and
anharmonic phonon renormalization as coupled microscopic routes for tuning phonon-mediated superconductivity.

Given the central role of Nb in superconducting radio-frequency cavities and quantum devices, where surface preparation, crystallographic quality, and residual strain are key experimental handles, the microscopic picture developed here may inform strategies for optimizing superconducting properties and potentially mitigating quasiparticle losses in these applications.


\begin{acknowledgments}
This work was supported by the Enterprise Science Fund of Intellectual Ventures and the Defense Advanced Research Projects Agency (DARPA), Defense Sciences Office (DSO), under grant HR0011-24-9-0394. Computations were performed on the lCluster at TU Graz and the Austrian Scientific Computing clusters VSC4 and VSC5. PNF acknowledges support from the Austrian Science Fund (FWF) under project DOI 10.55776/ESP8588124.
\end{acknowledgments}
\vspace*{1em}

\appendix

\section{Computational details}
\label{app:compdetails}
 
\textit{DFT parameters.}
A plane-wave kinetic energy cutoff of 140~Ry was used for all calculations. Brillouin-zone sampling for bulk Nb employed a
$24\times24\times24$ Monkhorst--Pack \textbf{k}-point mesh~\cite{monkhorst1976}; for the (001), (110), and (111) surface
slabs, meshes of $18\times18\times1$, $12\times16\times1$, and $12\times12\times1$ were used, respectively. Electronic occupations were treated with the Methfessel--Paxton smearing scheme~\cite{MP-smearing} at a broadening of 0.02~Ry. Structural relaxation was performed until total energies and atomic forces converged to $1\times10^{-7}$~Ry and $1\times10^{-6}$~Ry/Bohr, respectively. Slabs were built from the optimized bulk geometry using the VASPKIT program~\cite{VASPKIT}; the (001), (110), and (111) supercells contain 6, 12, and 12 Nb atoms, respectively.
 
\textit{DFPT and mode-resolved EPC.}
Phonon Brillouin zones were sampled on uniform \textbf{q}-point meshes of $6\times6\times6$ (bulk), $8\times8\times1$ (001), $4\times6\times1$ (110), and $4\times4\times1$ (111), from which the DFPT phonon linewidths entering Eq.~(1) were evaluated. For the fat-band visualization of mode- and wavevector-resolved EPC shown in Figs.~\ref{fig:fig2_bulk_strain_main}(d) and (e),
EPC matrix elements were Wannier-interpolated using the \textsc{EPW} code onto dense \textbf{k}-point grids  $60\times60\times60$ for bulk; analogous two-dimensional grids for the slabs, with a single point along the non-periodic direction.
 
\textit{SSCHA.}
For the slabs, minimizations were performed in $2\times2\times1$ supercells at 5~K using up to 5000 MLIP-evaluated configurations per minimization step, and were considered converged when the ratio of the free-energy gradient with respect to the auxiliary dynamical matrix to its stochastic error fell below $10^{-7}$. The SSCHA free-energy Hessian was Fourier-interpolated to obtain the anharmonically renormalized phonon dispersions shown in Fig.~\ref{fig:fig4_surface_anharm}. For bulk Nb, SSCHA minimizations were performed at 5~K in $2\times2\times2$ supercells using the same MLIP and convergence criteria. Figures~\ref{fig:fig9_MLIP_force_rmse} and~\ref{fig:fig10_MLIP_energy_rmse} report the force and energy validation of the trained MLIP against independent DFT test configurations for each surface termination. To investigate the sensitivity of anharmonic frequencies to the temperature, additional minimizations were performed at 10~K and 15~K for bulk Nb at all four lattice constants, and at 10~K for the three slabs using similar input parameters.

\textit{Numerical convergence.}
Figures~\ref{fig:convergence}(a) and (c) show \tc{} and $\lambda$, respectively, as a function of the $k$-point mesh ($12^3$--$24^3$) for plane-wave cutoffs between 80 and 140~Ry at a fixed $q$-mesh of $6\times6\times6$. The calculated \tc{} varies only between 9.0 and 9.5~K across all tested meshes and cutoffs, while $\lambda$ varies only moderately. At the $24\times24\times24$ $k$-point mesh, the different cutoff energies give $\lambda=0.90$--0.97 and $T_\text{c}=9.0$--9.5~K. Across the full range of tested $k$-point meshes, the variation remains below approximately 0.6~K in \tc{} and 0.07 in $\lambda$. We additionally tested a higher plane-wave cutoff of 160~Ry using a
$24\times24\times24$ $k$-point mesh, obtaining $\lambda=0.914$ and $T_\text{c}=9.4$~K. Furthermore, increasing the $k$-point mesh from $24\times24\times24$ to $28\times28\times28$ at $E_{\mathrm{cut}}=140$~Ry gives $\lambda=0.97$,  $\omega_{\log}=12.82$~meV, and $T_\text{c}=9.5$~K. These additional tests show that the superconducting transition temperature is essentially unchanged upon further increasing either the plane-wave cutoff or the $k$-point density, supporting our choice of $E_{\mathrm{cut}}=140$~Ry and a $24\times24\times24$ $k$-point mesh.

Figures~\ref{fig:convergence}(b,d) show the corresponding convergence with respect to the phonon $q$-point mesh ($2^3$--$10^3$), using $E_{\mathrm{cut}}=140$~Ry and a $24\times24\times24$ $k$-point mesh. From $6\times6\times6$ onward, \tc{} and $\lambda$ show only small variations; increasing the mesh from $6\times6\times6$ to $8\times8\times8$ and $10\times10\times10$ changes \tc{} by less than 0.2~K and $\lambda$ by less than 0.03. These results demonstrate that the calculated superconducting transition temperature is well converged with respect to the $q$-point sampling.

For the Nb surface slabs, the DFT/DFPT calculations employ the same plane-wave cutoff, pseudopotentials, and numerical thresholds validated by the bulk convergence tests. The differences in \tc{} between the three terminations (at least 1.6~K) are more than twice as large as the variations found in these tests (below 0.6~K for the $k$-point and below 0.2~K for the $q$-point sampling). We therefore expect the orientation dependence of \tc{} to be robust with respect to the Brillouin-zone sampling.

\section{Thickness and strain dependence of Nb slabs}
\label{app:strain-thick-slab}

\subsection{Thickness dependence}
\begin{figure}[htbp]
    \centering
    \includegraphics[width=0.5\textwidth]{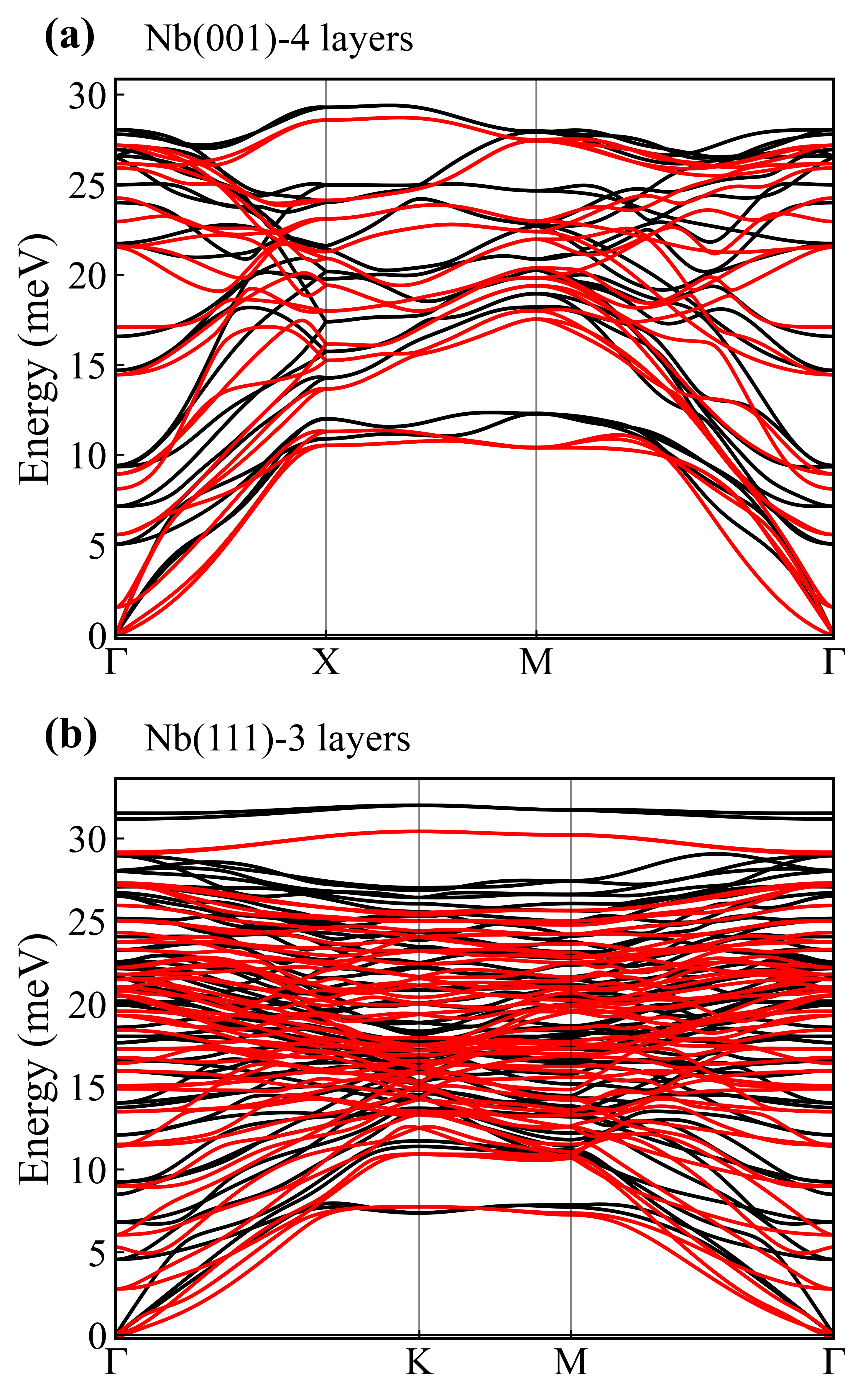}
    \caption{Harmonic (black) and anharmonic (red) phonon dispersions of (a) four-layer Nb(001) and (b) three-layer Nb(111) slabs. The anharmonic phonon dispersions are obtained from SSCHA calculations at 5~K.}
    \label{thickness-phonons}
\end{figure}

A fully first-principles thickness-dependent study of larger slabs using DFPT is computationally very demanding, particularly when anharmonic lattice effects are included. Nevertheless, it is important to evaluate whether finite slab thickness introduces qualitative features that disappear in the bulk limit. We therefore performed additional thickness-dependent calculations for Nb(001), comparing three- and four-layer slabs ($\sim 11.2$~\AA{} thick), and for Nb(111), comparing two- and three-layer slabs ($\sim 15.0$~\AA{} thick), using both harmonic and anharmonic phonon calculations. The phonon dispersions of the thicker slabs are shown in Fig.~\ref{thickness-phonons}, and the calculated superconducting parameters are summarized in Table~\ref{tab:thickness-Nb001-Nb111}.

\begin{table}[hbtp]
\centering
\caption{Superconducting properties of Nb(001) and Nb(111) slabs with different thicknesses. The anharmonic results are obtained from SSCHA calculations at 5~K.}
\label{tab:thickness-Nb001-Nb111}
\setlength{\tabcolsep}{5pt}
\begin{tabular}{c c c c c}
\hline\hline
Slab & Method & $\lambda$ &
$\omega_{\log}$ (meV) &
\tc{} (K) \\
\hline
Nb(001), 3 layers & SSCHA
& 1.06 & 11.90 & 10.1 \\

Nb(001), 4 layers & Harmonic
& 0.84 & 14.32 & 7.8 \\

Nb(001), 4 layers & SSCHA
& 0.97 & 12.45 & 9.2 \\
\hline
Nb(111), 2 layers & SSCHA
& 0.77 & 13.99 & 6.2 \\

Nb(111), 3 layers & Harmonic
& 0.74 & 15.31 & 6.1 \\

Nb(111), 3 layers & SSCHA
& 0.86 & 13.59 & 7.7 \\
\hline\hline
\end{tabular}
\end{table}

Increasing the slab thickness modifies the phonon spectrum and the resulting superconducting properties, as expected when moving from a thin, surface-dominated system toward a more bulk-like environment. For Nb(001), the electron–phonon coupling of the four-layer slab is reduced compared with the three-layer system, accompanied by an increase in the logarithmic phonon frequency and a reduction in \tc{} from 10.1~K to 9.2~K, close to the bulk value. For Nb(111), the opposite trend is found: the electron–phonon coupling of the three-layer slab is enhanced compared with the two-layer system ($\lambda$ = 0.86 vs. 0.77), and \tc{} increases from 6.2~K to 7.7~K. Thus, with increasing thickness, the transition temperatures of both terminations move toward the anharmonic bulk value of 8.5~K (Table~\ref{tab:anharmonic-bulk}) from opposite sides. The ordering $T_\text{c}(001)>T_\text{c}(111)$ is preserved, although the difference decreases from 3.9~K to 1.5~K. 

\begin{figure*}[hbtp]
    \centering
    \includegraphics[width=1.0\textwidth]{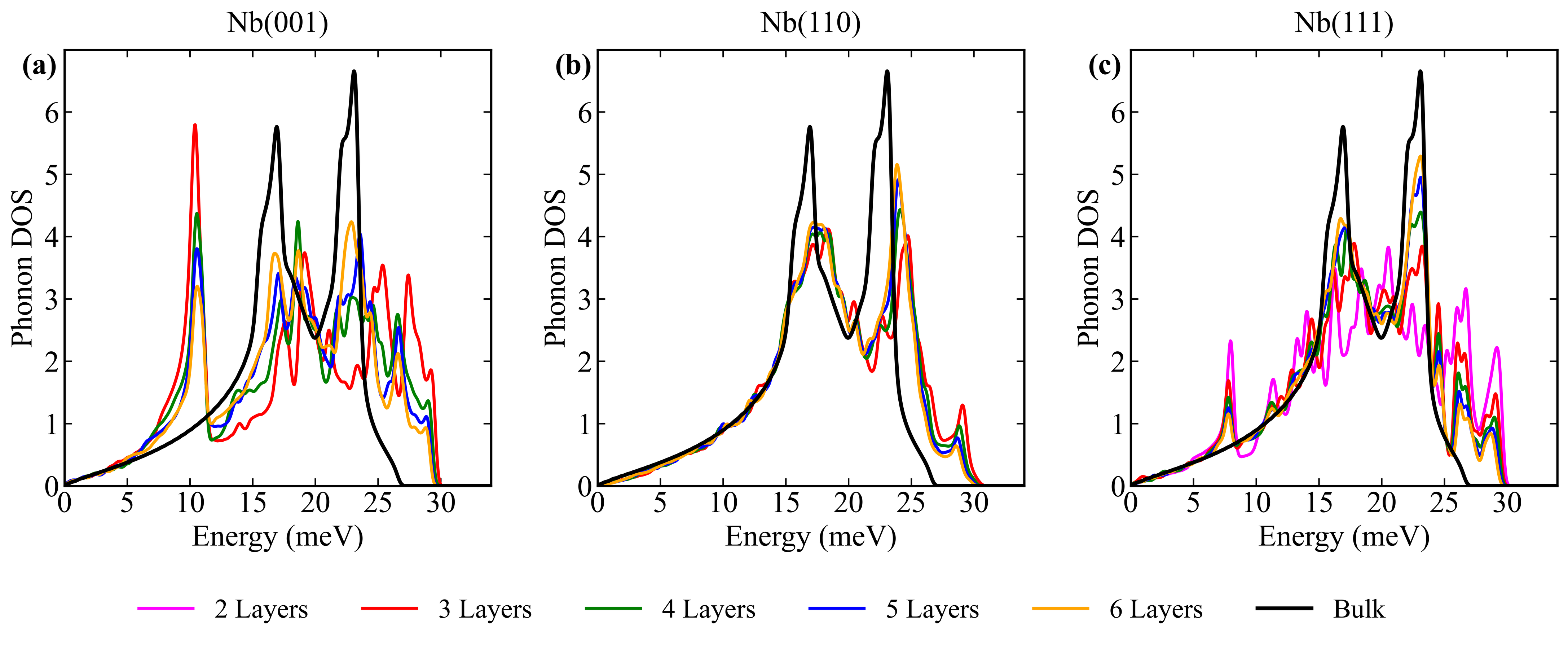}
    \caption{
    Normalized harmonic phonon density of states of Nb slabs with different 
    thicknesses for (a) Nb(001), (b) Nb(110), and (c) Nb(111) surfaces. The 
    phonon spectra were calculated using the machine-learning interatomic 
    potential combined with LAMMPS and Phonopy. }
    \label{fig:thickness-DOS}
\end{figure*}

To further examine the evolution of lattice dynamics with increasing thickness, we calculated the harmonic phonon density of states for Nb(001) and Nb(110) slabs containing three to six layers and for Nb(111) slabs containing two to six layers. These calculations were carried out using our Nb machine-learning interatomic potential combined with LAMMPS and Phonopy~\cite{novikov2021mlip,THOMPSON2022108171,TOGO20151}, which allows an efficient comparison of multiple surface orientations and slab thicknesses. The resulting phonon DOS are compared with the bulk Nb phonon DOS in Fig.~\ref{fig:thickness-DOS}.

The thickness dependence reveals a gradual evolution of the surface vibrational spectrum toward the bulk limit. For all three surface orientations, thinner slabs exhibit additional surface-related vibrational features and deviations from the bulk phonon spectrum. With increasing slab thickness, these features are progressively reduced, and the overall spectral distribution becomes more similar to the bulk phonon DOS. The rate of this evolution depends on the surface orientation. Nb(110) shows relatively small thickness dependence, whereas Nb(001) and Nb(111) exhibit stronger modifications, particularly in the low-frequency region. This trend is physically relevant for superconductivity, because low-frequency phonon modes generally provide an important contribution to the electron–phonon coupling. The gradual suppression of surface-related low-frequency vibrational features with increasing thickness suggests that the surface contribution to the electron–phonon interaction decreases as the slab approaches a more bulk-like environment.

We emphasize, however, that the phonon density of states alone does not establish quantitative convergence of $\lambda$ or \tc{}. A complete thickness-dependent investigation would require explicit DFPT calculations for each slab thickness and orientation, which become computationally prohibitive for larger slabs. The present analysis therefore provides qualitative insight into the evolution of surface lattice dynamics and supports the interpretation that surface-related superconducting contributions gradually evolve toward the bulk limit with increasing thickness.

\subsection{Epitaxial strain on Nb(001)}
\begin{figure*}[hbpt]
    \centering
    \includegraphics[width=0.95\textwidth]{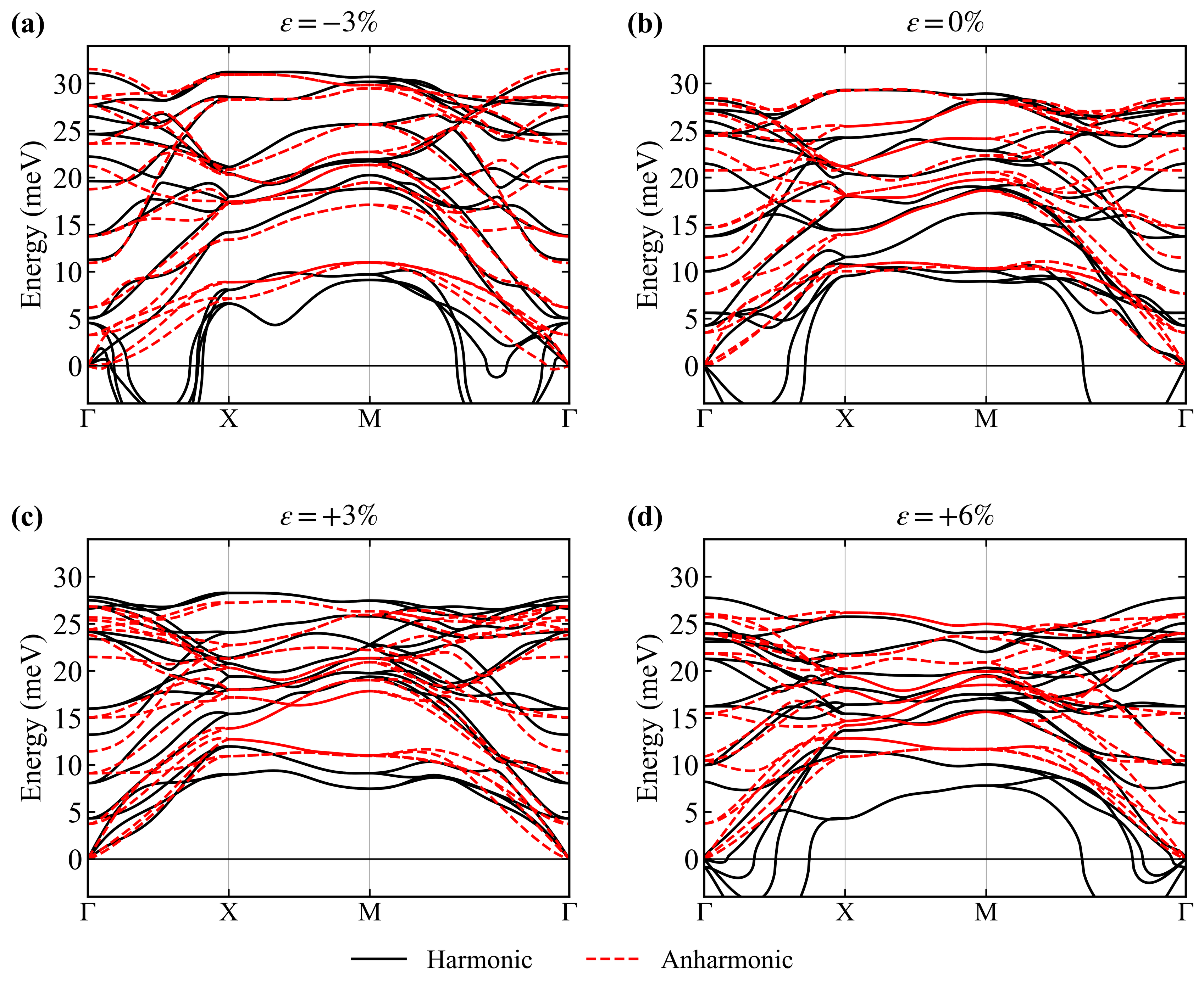}
    \caption{Harmonic and anharmonic phonon dispersions of the three-layer Nb(001) slab under epitaxial strain. Panels (a)--(d) correspond to compressive ($\epsilon=-3\%$), equilibrium ($\epsilon=0\%$), and tensile ($\epsilon=+3\%$ and $+6\%$) strain, respectively. Black solid lines denote harmonic phonon spectra, while red dashed lines represent the anharmonic phonon dispersions obtained from SSCHA calculations at 5~K.}
    \label{fig:Nb001-strain-phonon}
\end{figure*}

In experimentally fabricated thin films, substrate-induced strain is commonly present. To examine its effect on the slab results, we focus on the Nb(001) surface, which exhibits the highest superconducting transition temperature among the three investigated orientations. We performed additional calculations for the three-layer slab under in-plane compressive ($-3\%$) and tensile ($+3\%$ and $+6\%$) strain relative to the equilibrium slab. The harmonic and anharmonic phonon dispersions are shown in Fig.~\ref{fig:Nb001-strain-phonon}, and the corresponding superconducting parameters are summarized in Table~\ref{tab:surface-strain}.

As in the unstrained slab, the harmonic phonons of the $-3\%$ and $+6\%$ structures contain imaginary modes, so their superconducting properties were evaluated using the anharmonically renormalized SSCHA spectra, which are dynamically stable in all cases. The $+3\%$ structure, in contrast, is harmonically stable, allowing a direct comparison between harmonic and anharmonic results.

Interestingly, the strain dependence of the superconducting properties of Nb(001) is opposite to that of bulk Nb. Compressive strain softens the low-energy surface modes that dominate the coupling, enhancing the electron--phonon coupling and raising \tc{} to approximately $14$~K, compared with $10.1$~K for the unstrained slab. Tensile strain instead stiffens these modes, reducing the coupling and progressively lowering \tc{} with increasing strain. At $+3\%$ strain, anharmonic renormalization further reduces the coupling relative to the harmonic result and lowers \tc{} by about $1$~K. These results demonstrate that epitaxial strain can significantly modify the superconducting properties of Nb surfaces and that anharmonic phonon effects remain important under strained conditions.

\begin{table}[h]
\centering
\caption{Superconducting properties of the three-layer Nb(001) slab under epitaxial strain. The anharmonic results are obtained from SSCHA calculations at 5~K. For the $+3\%$ tensile-strained structure, both harmonic and anharmonic results are compared.}
\label{tab:surface-strain}
\begin{tabular}{cccccc}
\hline\hline
$\epsilon$ & $a$ (\AA) & Method & $\lambda$ & $\omega_{\log}$ (meV) 
& \tc{} (K)\\
\hline
$-3\%$ & 3.17 & SSCHA & 1.66 & 8.91  & 14.0\\
$0\%$  & 3.27 & SSCHA & 1.06 & 11.93 & 10.1\\
$+3\%$ & 3.37 & Harmonic & 1.00 & 11.32  & 8.7\\
$+3\%$ & 3.37 & SSCHA & 0.87 & 13.11  & 7.7\\
$+6\%$ & 3.47 & SSCHA & 0.81 & 13.35 & 6.7 \\
\hline\hline
\end{tabular}
\end{table}

\section{Additional Figures and Tables}
\label{app:addfigures}

\begin{figure*}[t]
    \centering
    \includegraphics[width=1\linewidth]{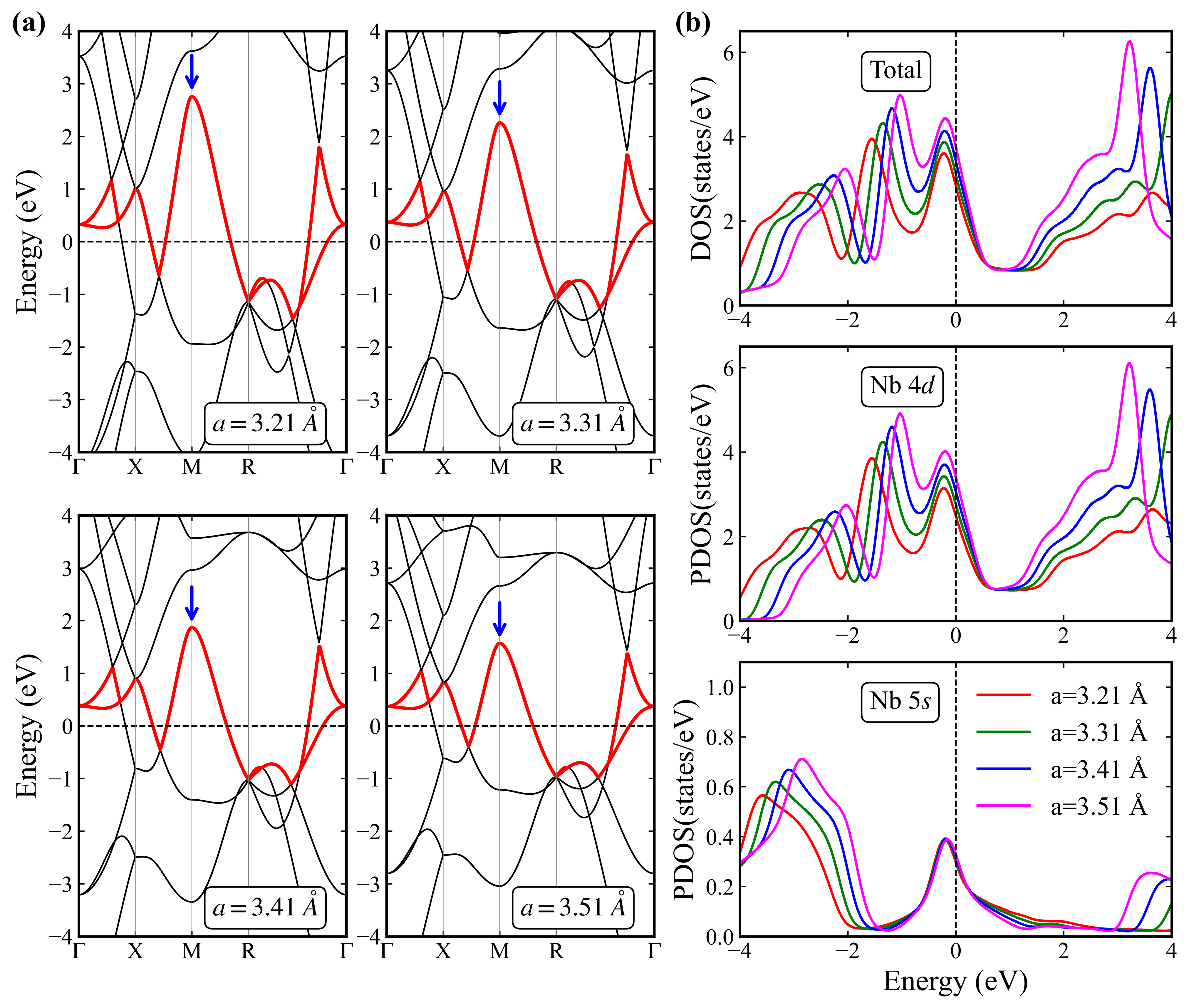}
    \caption{(a) Electronic band structures of bulk Nb at four lattice constants along the high-symmetry path $\Gamma$--X--M--R--$\Gamma$. The electronic states highlighted in red mark the region near the Fermi level whose evolution with strain is discussed in the text; the blue arrow indicates the local band maximum in this region. The dashed horizontal line marks the Fermi level ($E = 0$). (b) Total DOS and projected DOS onto the Nb~$4d$ and $5s$ orbitals for the same four lattice constants (color coding as indicated in the legend). The dashed vertical line marks the Fermi level.}  
    \label{fig:fig7_bulk_elbands_dos}
\end{figure*}

\begin{figure*}[hbtp]
    \centering
    \includegraphics[width=0.8\textwidth]{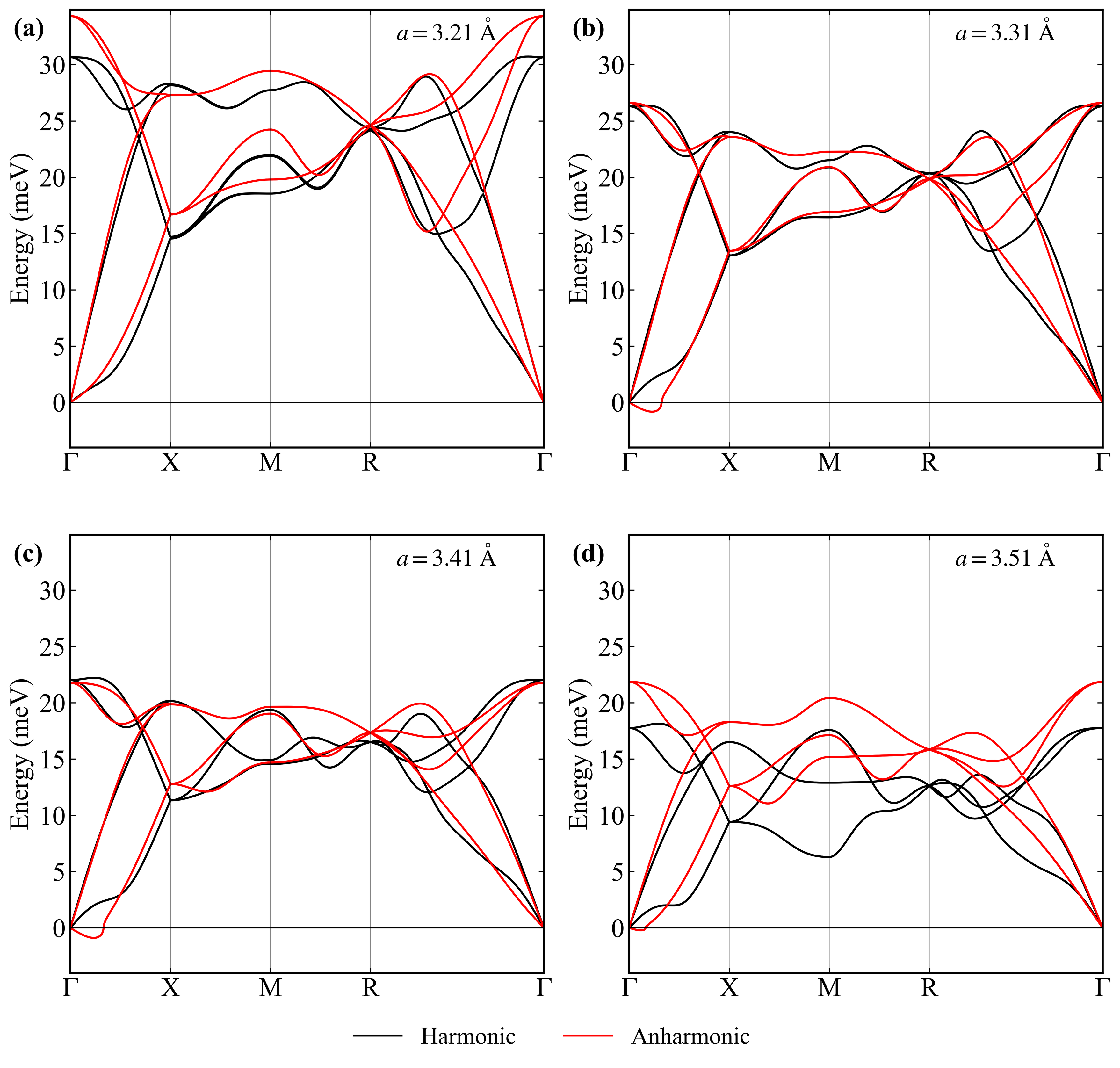}
    \caption{%
    Harmonic (black) and anharmonic (red) phonon dispersions of bulk Nb at different lattice parameters: (a) $a=3.21$~\AA, (b) $a=3.31$~\AA, (c) $a=3.41$~\AA, and (d) $a=3.51$~\AA. The anharmonic phonon spectra are obtained from SSCHA calculations at 5~K. The small negative frequencies of the SSCHA spectra close to $\Gamma$ are an artifact of the Fourier interpolation from $2\times2\times2$ supercells.}
    \label{fig:bulk-phon-har-anhar}
\end{figure*}

\begin{figure*}[hbtp]
    \centering
    \includegraphics[width=0.8\textwidth]{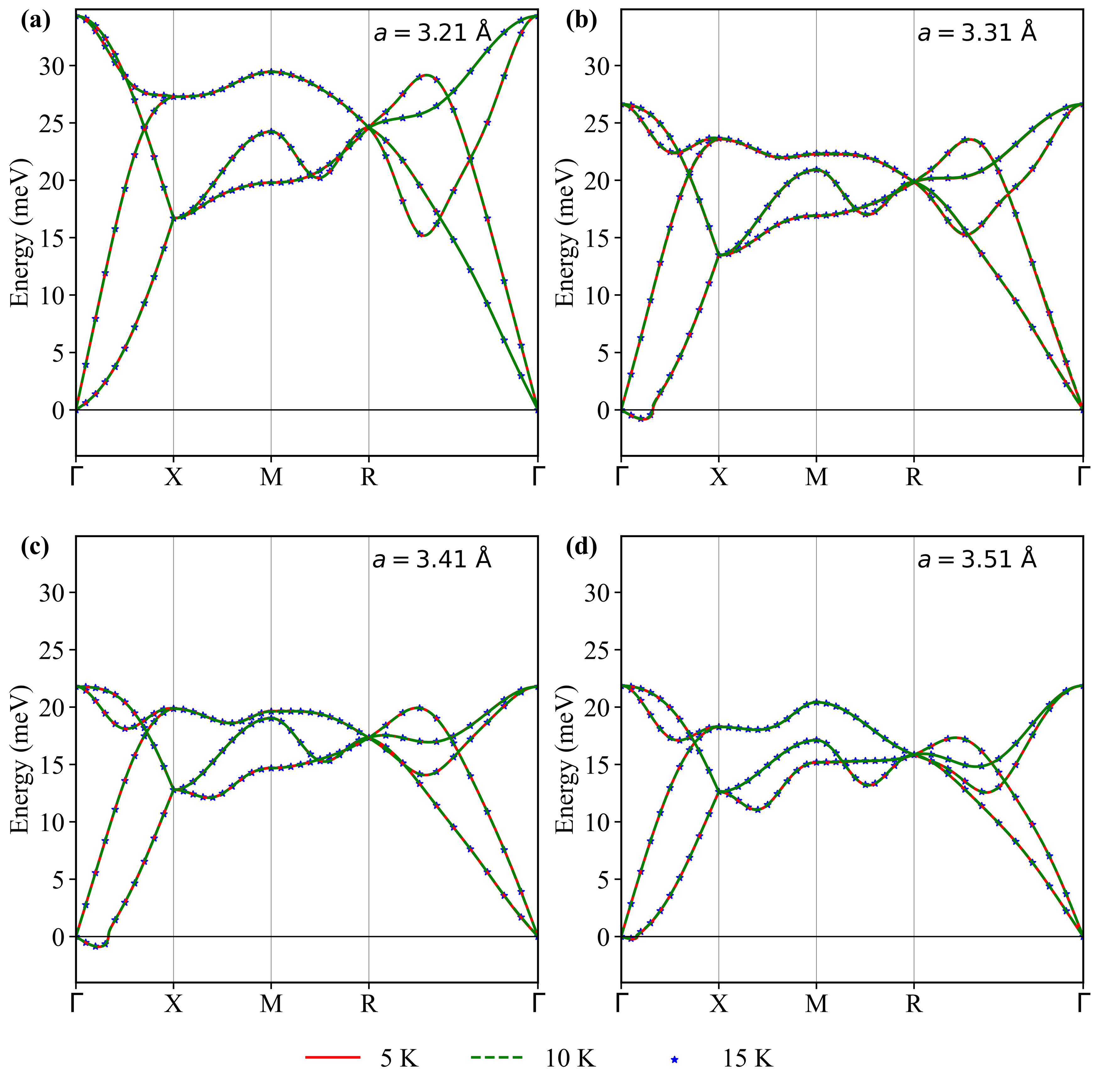}
    \caption{%
    Temperature dependence of the anharmonic phonon dispersions of bulk Nb for lattice parameters (a) $a=3.21$~\AA, (b) $a=3.31$~\AA, (c) $a=3.41$~\AA, and (d) $a=3.51$~\AA. The phonon dispersions calculated at 5~K (red solid lines), 10~K (green dashed lines), and 15~K (blue stars) show negligible temperature dependence.}
    \label{fig:bulk-phon-temp}
\end{figure*}

\begin{figure*}[hbtp]
    \centering
    \includegraphics[width=\textwidth]{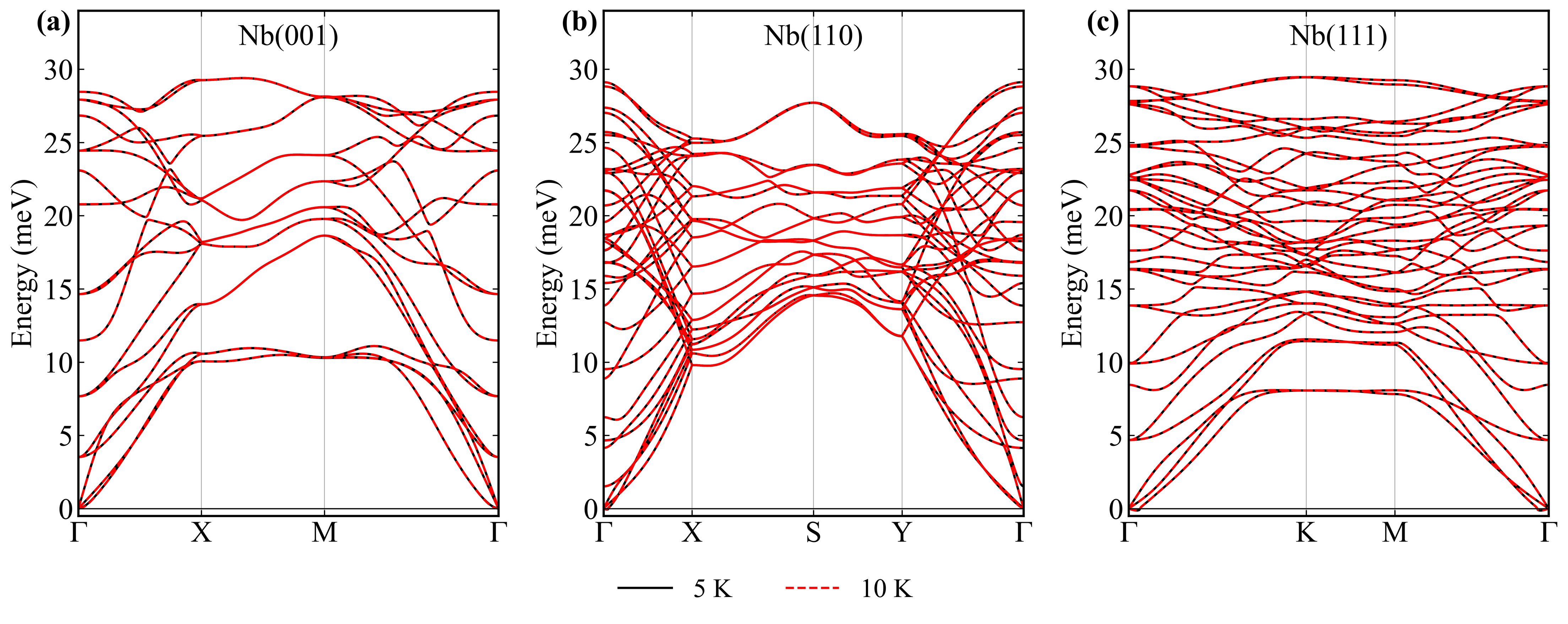}
    \caption{%
    Temperature dependence of the anharmonic phonon dispersions for Nb slabs: (a) three-layer Nb(001), (b) three-layer Nb(110), and (c) two-layer Nb(111). The phonon dispersions calculated at 5~K (black solid lines) and 10~K (red dashed lines) are nearly identical, indicating negligible temperature-induced changes in the anharmonic phonon spectrum.}
    \label{fig:surf-phon-temp}
\end{figure*}

\begin{figure*}[t]
    \centering
    \includegraphics[width=1.0\linewidth]{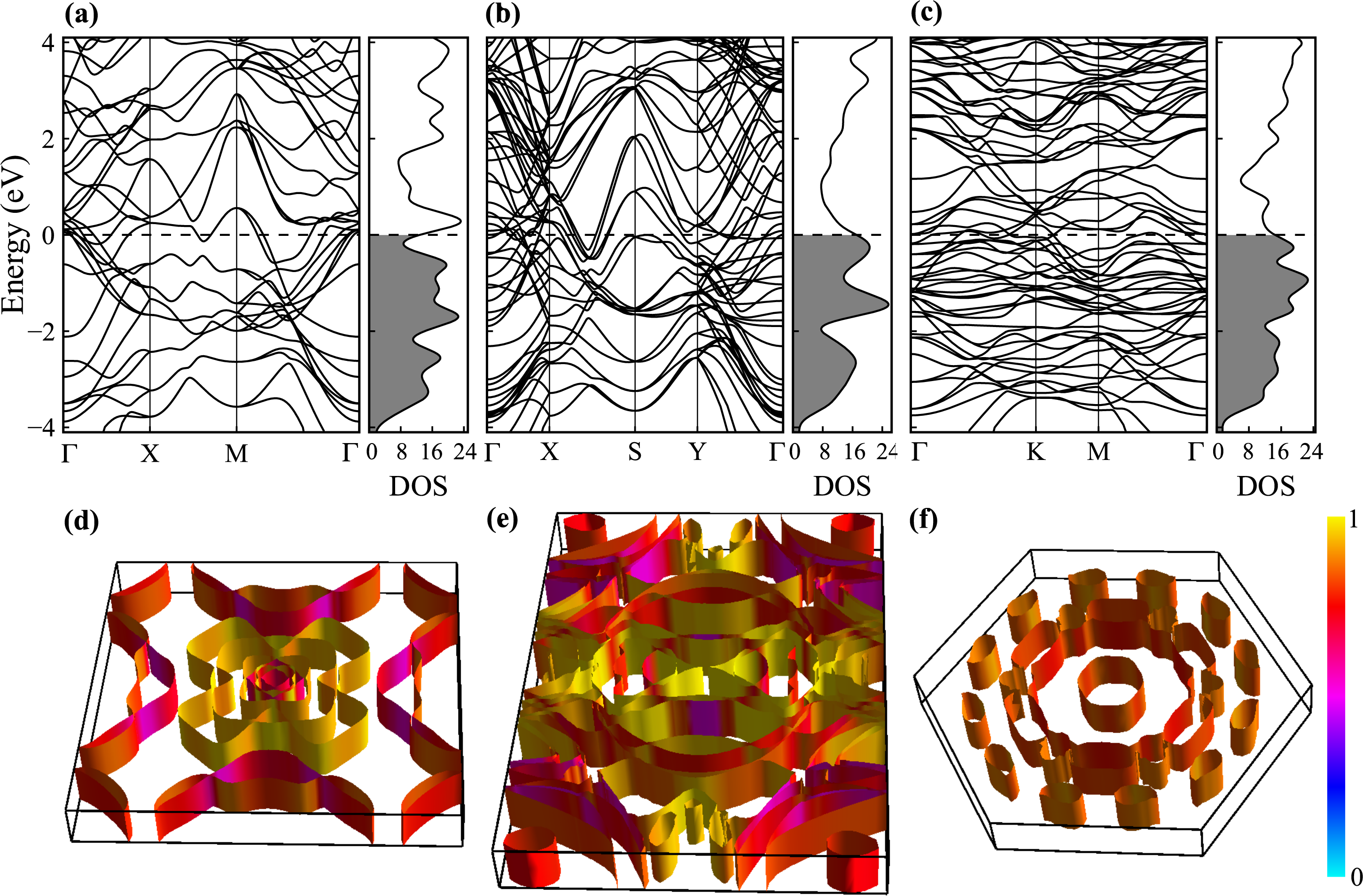}
    \caption{Electronic band structures with corresponding total DOS (states/eV) and Fermi surfaces of the three Nb slabs. (a)--(c) Band structures of the Nb(001), Nb(110), and Nb(111) slabs along their respective two-dimensional high-symmetry paths, with the total DOS shown to the right of each panel. The dashed  horizontal line marks the Fermi level ($E = 0$). (d)--(f) Corresponding two-dimensional Fermi surfaces colored by the projected Nb $4d$ orbital weight (color scale shown on the right).}
    \label{fig:fig8_slabs_elbands_dos_FS}
\end{figure*}

\begin{figure*}[t]
    \centering
    \includegraphics[width=1\linewidth]{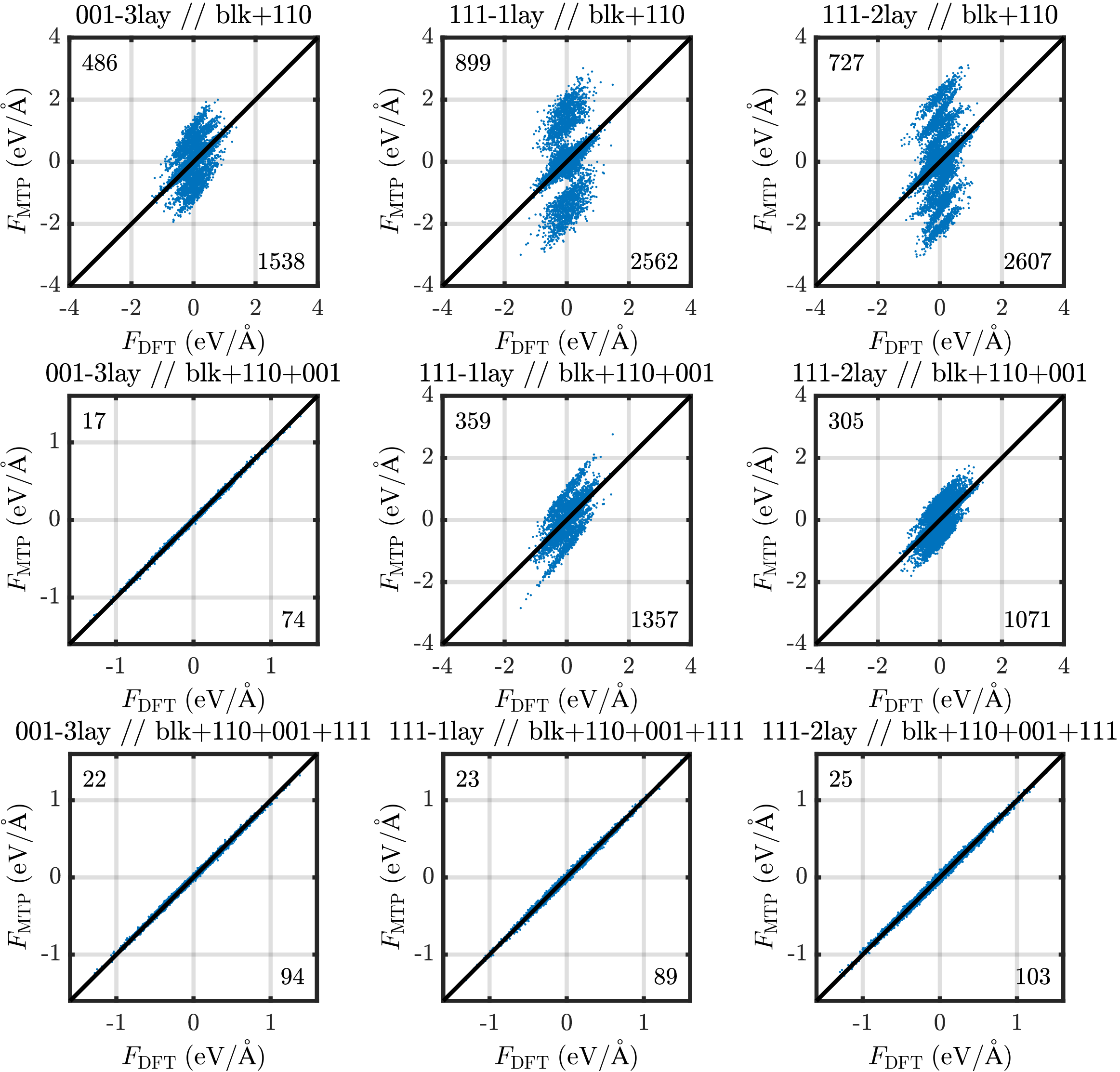}
    \caption{Parity plots of MLIP-predicted versus DFT atomic forces for three validation structures (columns: 001-3lay, 111-1lay, 111-2lay) and three training set compositions (rows: Bulk$+$Nb(110), Bulk$+$Nb(110)$+$Nb(001), and Bulk$+$Nb(110)$+$Nb(001)$+$Nb(111)). The root-mean-square error and maximum absolute deviation of the predicted forces, both in meV/\AA, are given in the top-left and bottom-right corners of each panel, respectively.}
    \label{fig:fig9_MLIP_force_rmse}
\end{figure*}

\begin{figure*}[t]
    \centering
    \includegraphics[width=1\linewidth]{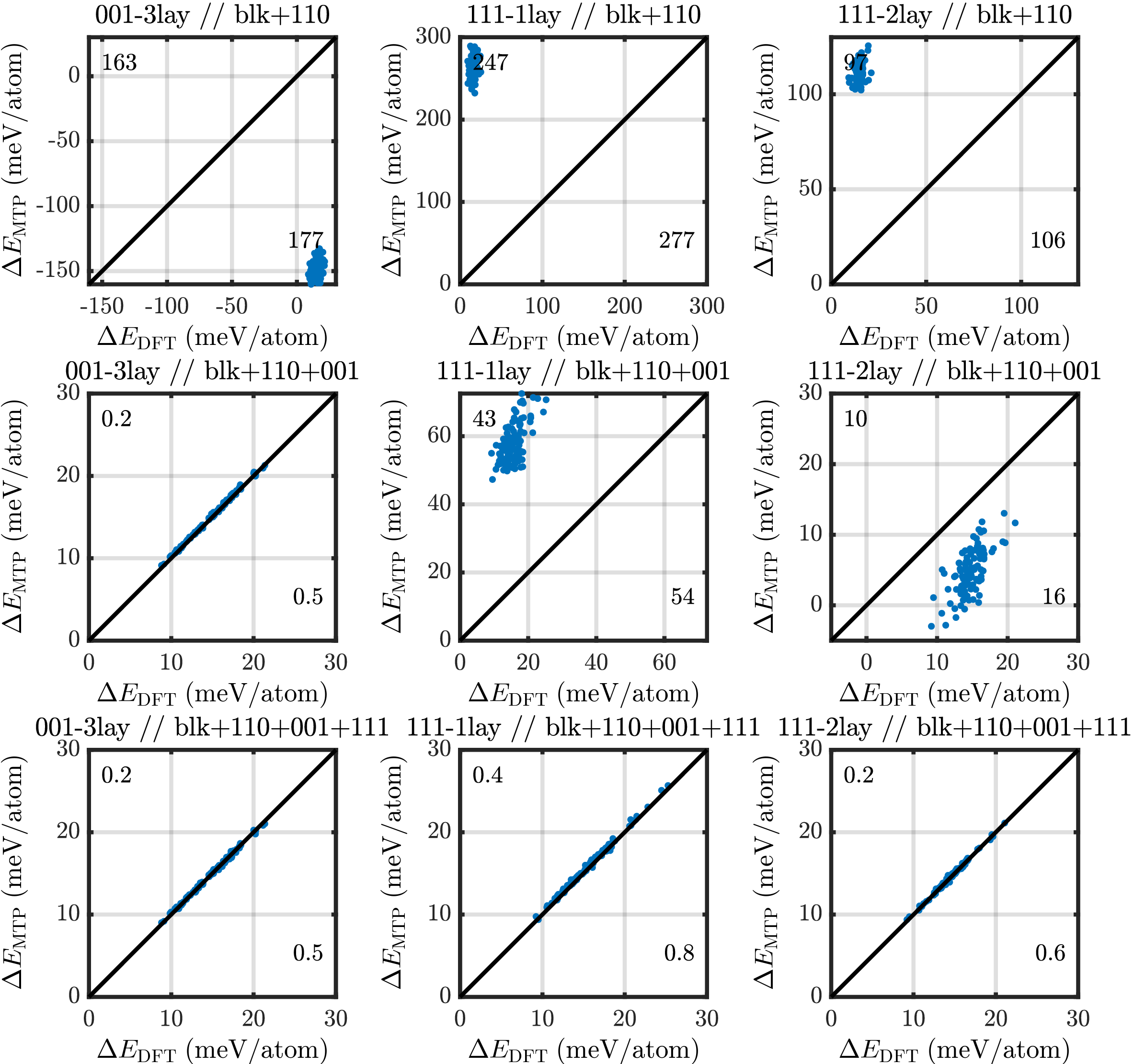}
    \caption{Parity plots of MLIP-predicted versus DFT formation energies for the same validation structures and training set compositions as in Fig.~\ref{fig:fig9_MLIP_force_rmse}. The root-mean-square error and maximum absolute deviation, both in meV/atom, are given in the top-left and bottom-right corners of each panel, respectively.} 
    \label{fig:fig10_MLIP_energy_rmse}
\end{figure*}

\begin{figure*}[hbtp]
    \centering
    \includegraphics[width=\textwidth]{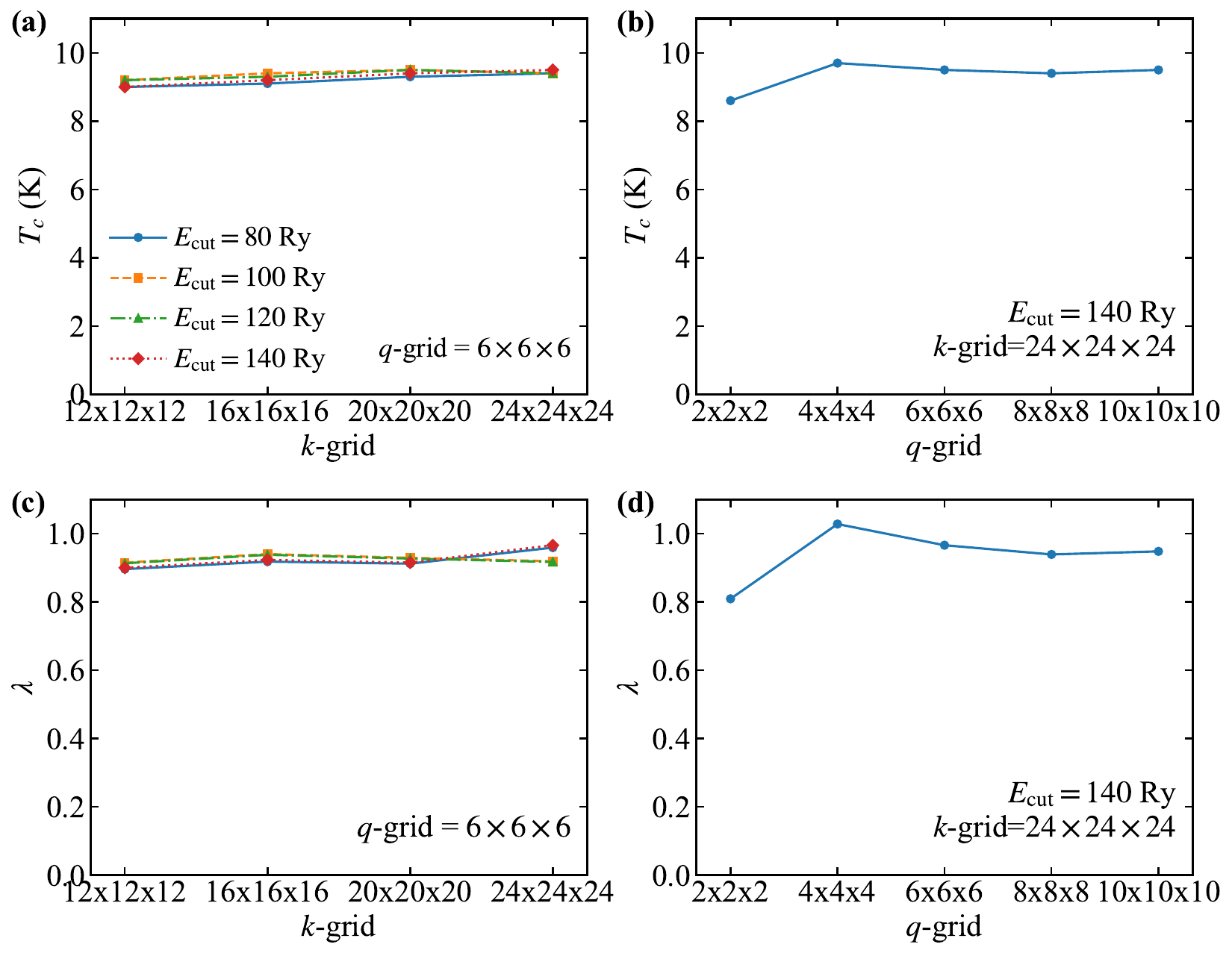}
    \caption{%
    Convergence of the superconducting transition temperature \tc{} and
    electron--phonon coupling constant $\lambda$ with respect to
    Brillouin-zone sampling for bulk Nb.
    (a,c) \tc{} and $\lambda$, respectively, as a function of the
    $k$-point grid for different plane-wave energy cutoffs
    $E_{\mathrm{cut}}$, with the $q$-point grid fixed at
    $6\times6\times6$.
    (b,d) \tc{} and $\lambda$, respectively, as a function of the
    $q$-point grid, using $E_{\mathrm{cut}}=140$ Ry and a
    $24\times24\times24$ $k$-point grid.
    }
    \label{fig:convergence}
\end{figure*}

\begin{table*}[hbtp]
\centering
\caption{Sensitivity of the calculated superconducting transition temperature \tc{} to the Coulomb interaction for bulk Nb at different lattice parameters and for the Nb(001), Nb(110), and Nb(111) slabs. The density of states at the Fermi level, $N(E_F)$, is given per two-atom cubic cell.}
\label{tab:mu-sensitivity}
\small
\setlength{\tabcolsep}{5pt}
\begin{tabular}{l c c c c c c}
\hline\hline
System &
\makecell{$N(E_F)$\\(states/eV/spin)} &
$\mu_{\mathrm{new}}$ &
\makecell{\tc{} ($\mu=0.38$)\\(K)} &
\makecell{\tc{} ($\mu_{\mathrm{new}}$)\\(K)} &
\makecell{\tc{} ($\mu^*_{\mathrm{AD}}=0.10$)\\(K)} &
\makecell{\tc{} ($\mu^*_{\mathrm{AD}}=0.16$)\\(K)}
\\
\hline
Bulk 3.21~\AA\ ($-3\%$) & 1.27 & 0.34 & 8.2 & 8.4 & 9.4 & 7.0 \\
Bulk 3.31~\AA\ (eq.)   & 1.40 & 0.38 & 9.5 & 9.5 & 10.5 & 8.8 \\
Bulk 3.41~\AA\ ($+3\%$) & 1.53 & 0.42 & 11.3 & 11.1 & 12.1 & 10.6 \\
Bulk 3.51~\AA\ ($+6\%$) & 1.67 & 0.46 & 14.6 & 14.3 & 15.2 & 14.0 \\
\hline
Nb(001) & 1.26$^{*}$ & 0.34 & 10.1 & 10.3 & 11.1 & 9.5 \\
Nb(110) & 1.32$^{*}$ & 0.36 & 7.8 & 7.9 & 8.8 & 4.7 \\
Nb(111) & 1.22$^{*}$ & 0.33 & 6.2 & 6.5 & 7.2 & 3.4 \\
\hline\hline
\end{tabular}

\vspace{2pt}
\begin{flushleft}
\footnotesize
$^{*}$ For the surface slabs, $N(E_F)$ is normalized to the two-atom bulk cell by dividing the slab values by 3, 6, and 6 for Nb(001), Nb(110), and Nb(111), respectively.
\end{flushleft}
\end{table*}

\bibliography{reference}

\end{document}